\documentclass[fleqn,usenatbib]{mnras}

\usepackage{newtxtext,newtxmath}

\usepackage[T1]{fontenc}

\DeclareRobustCommand{\VAN}[3]{#2}
\let\VANthebibliography\thebibliography
\def\thebibliography{\DeclareRobustCommand{\VAN}[3]{##3}\VANthebibliography}

\usepackage{graphicx}	
\usepackage{amsmath}	
\usepackage{subcaption}
\usepackage{siunitx}
\usepackage[normalem]{ulem}
\usepackage{gensymb}

\title[Migration in fragmenting MHD discs]{The stochastic nature of migration of disc instability protoplanets in three-dimensional hydrodynamical and MHD simulations of fragmenting discs}

\author[Kubli et al.]{
Noah Kubli,$^{1}$\thanks{E-mail: noah.kubli@uzh.ch}
Lucio Mayer,$^{1}$
Hongping Deng,$^{2}$
Douglas N. C. Lin$^{3,4}$
\\
$^{1}$ Department of Astrophysics, University of Zurich, Winterthurerstrasse 190, CH-8057 Zürich, Switzerland
\\
$^{2}$ Shanghai Astronomical Observatory, Chinese academy of Science, Nandan Rd 80th, 200030 Shanghai, China\\
$^{3}$ Department of Astronomy and Astrophysics, University of California, Santa Cruz, CA 95064, USA\\
$^{4}$ Institute for Advanced Studies, Tsinghua University, Beijing 100084, China
}

\date{Accepted XXX. Received YYY; in original form ZZZ}

\pubyear{\the\year{}}

\begin{document}
\label{firstpage}
\pagerange{\pageref{firstpage}--\pageref{lastpage}}
\maketitle

\begin{abstract}
We present a detailed analysis of the nature of migration of protoplanetary clumps formed via disc instability in self-consistent 3D 
hydrodynamical (HD) and magneto-hydrodynamical (MHD) simulations of self-gravitating discs. Motivated by the complex structure of protoplanetary clumps we do not introduce sink particles.
We find that the orbital evolution of
the clumps has a stochastic character but also exhibits
recurrent properties over many orbits.
Clump migration is governed by two sources of gravitational torques: a torque originating from a region about twice the Hill sphere around each clump's orbit, and the torque resulting from clump-clump interactions. 
Compared to non-magnetized companion runs, the latter are more frequent in MHD simulations, which give rise to more numerous clumps starting off at smaller masses,
often below a Neptune mass.
Clump-clump interactions can lead to temporary strong accelerations
of migration in both directions,
but integrated over time provide a lesser impact
than disc-driven torques. 
They can also lead to clump mergers but do not cause ejections;
a difference to previous works which adopted sink particles. 
The local ``Hill torque'' is responsible
for the fast migration, inward or outward. 
Estimating the characteristic timescales of conventional migration in our regime,
we find that the disc-driven migration timescales are in agreement with Type III migration. 
However, the dominant
local torque is rapidly fluctuating, which reflects the turbulent nature of the flow.
The resulting stochastic migration pattern
is markedly different from Type III runaway migration and appears
to be a distinctive feature of orbital dynamics in a fragmenting disc.
\end{abstract}




\section{Introduction}
Gravitational instability (GI) \citep{kuiper}
is a possible formation path of planets.
For sufficiently cold gas, gravity
may dominate over shear and pressure forces 
leading to a local collapse in the
protoplanetary disc \citep{toomre, boss-97, mayer-2002}.
Although conventional wisdom attributes a lesser
role of disc instability theory relative to the core
accretion (CA) theory \citep{safronov-72, pollack-92, johansen-17} in the process of planet formation,
the relative importance of the two modes of planet
formation is still not clear.
For very massive planets $> 4 M_\text{J}$
the correlation between metallicity and giant planet
occurrence observed at lower masses is no longer present,
disfavouring CA as the formation path for these planets
\citep{santos}.
Furthermore, CA has problems in explaining the observed vast
population of intermediate-mass planets \citep{suzuki, schlecker}.
Finally, the wide-orbit exoplanet population discovered by imaging
surveys
is becoming more numerous as time progresses
and is hosted by 
a wide range of stellar types 
\citep{marois-hr8799-08, bohn-20, janson-wideorbit-21, delorme-24}.
While this is not necessarily incompatible with
CA, it is more naturally explained by GI.
In both formation scenarios, planets are thought
to migrate radially 
\citep{goldreich-80, vorobyov-05, baruteau-11}
because of interactions with the disc in which they
are embedded as well as due to gravitational interactions with other planets.
Simple semi-analytical models of migration in GI have shown that it
can significantly affect the orbital distribution of planets after
birth, which, contrary to naive expectations, can result in a population
of Hot Jupiters as well as of wide-orbit planets \citep{galvagni-14, mueller-18}.

Global spiral structure in protoplanetary discs
in their early phases, the signpost of GI,
have been observed several times \citep{perez,meru,veronesi, speedie-24}
confirming that at least some discs may be 
gravitationally unstable \citep{toomre, deng-ogilvie-22}.
In a few cases even an analysis of the disc kinematics has been possible,
leading to results compatible with a self-gravitating disc
\citep{veronesi-24}.

Simulations of GI predict a wide range of fragment masses and radial 
separations from the host star,
from brown dwarfs and binary
stellar companions down to giant gas planets \citep{boss-97, mayer-2004}.
Recently, the first MHD simulations of GI showed fragmentation at much smaller
scales in the intermediate-mass
regime, generating Neptune-sized as well as Super-Earth-sized clumps \citep{deng-21, kubli}.
In the latter, magnetic fields stifle accretion via magnetic pressure, thus suppressing
mass growth after birth, and also shield them from disruption, thus allowing them to survive shear stresses and stellar tides despite their lower masses.
Magnetic fields in GI discs are amplified and sustained
through the spiral density waves by means of the so-called ``GI dynamo''
\citep{deng-20}.

Disc instability is favoured in  the early evolutionary stage 
of the disc, in Class 0-1, when it should be more massive  and
still accreting, two factors that can concur to trigger GI
and disc fragmentation 
\citep{boley-09, hayfield-11}.
It is therefore interesting to investigate the fate of
such clumps with respect to radial migration using simulations.
In the pure form of GI, the planet masses are correlated
with the distance to the central star. 
Similar to the CA model, migration and other processes, such as 
gas accretion \citep{zhu-12, stamatellos-15},
clump-clump collisions with 
various outcomes \citep{matzkevich-24},
pebble accretion \citep{humphries-18},
and tidal disruption and downsizing \citep{nayakshin-10} 
have to be taken into account
in order to explain the observed planetary 
population \citep{nayakshin-tdm3}.


In contrast to conventional migration theory \citep{goldreich-79, lin-79, tanaka-2002}, the presence of GI
adds additional effects:
the self-driven spiral patterns can induce torques on the (proto-)planets
leading to more chaotic dynamics
\citep{baruteau-11},
and clumps can interact with 
one another \citep{cha-2011, hall-17}.
Therefore simulations are needed to investigate
migration in such discs.
Migration in gravitationally unstable
discs has been investigated since more than a decade
using simulations.
Among the pioneering works is \citet{baruteau-11},
who
found inward migration on a time-scale
of 10 outer disc orbital periods using 2D simulations of 
individual protoplanets in a GI disc.
Other 2D simulations are used in \citet{machida-11} 
where they found 
infall of fragments on to the host star explaining
intermittent stellar outflows such as FU-Ori bursts.
\citet{malik-15} investigated the gap opening
criterion in both non-self-gravitating and GI discs, for both Type I and Type III migration of gas giants and brown dwarfs \citep{masset-03, masset-08, peplinski-12,baruteau-08, lin-papaloizou-12} and found that gap-opening is more difficult 
than previously thought
because a short migration time scale
compared to the orbital time
can prevent the opening of a gap where 
the conditions would otherwise be met.
In particular, in GI discs migration turned out to be often 
in the fast Type III regime, preventing gap formation.
On the other hand,
\citet{rowther-20}
found that migration can be
slowed down in GI discs once clumps enter the hot gravitationally
stable inner disc.
\citet{fletcher-19} carried out the first extensive code
comparison focused on migration of clumps that 
were formed in self-gravitating (SG)
discs. The GIZMO code \citep{gizmo1} we used in this paper was part of the
suit of employed codes. They found that 
the codes agreed
qualitatively yet the migration rate differed
by up to $50\%$.
To reduce complexity, they imposed a Toomre
$Q$ parameter near unity as expected in a GI
disc but without letting the disc fragment,
and inserted a planet represented by a sink particle 
with a mass and a formation location 
in the range expected for GI discs.

In most studies of migration in the disc instability 
scenario, as those just described, only a single clump is considered, using 
a sink particle 
(which can be accreting or not)
in a background disc that is marginally gravitationally
unstable but not fragmenting (e.g. \citet{baruteau-11, scott-11, malik-15, stamatellos-15}).
This is in contrast with the expected
configuration arising in fragmenting discs,
in which clumps are numerous and evolve in a disc
that has fragmented, which then
supports a highly gravito-turbulent flow \citep{durisen-07}.
In self-consistent simulations of fragmenting discs, the
clumps also mutually interact, can merge, and also 
accrete gas further while migrating.
In \citet{boss-12} and \citet{boss-23}, using two different Eulerian
hydrodynamics codes with flux-limited diffusion, 
they improved the realism of migration studies by 
simulating a disc in the 
fragmenting regime and inserting not one but
multiple clumps, represented as sink
particles, in the density maxima of spiral arms
where they expect fragmentation
to happen. In \citet{boss-23} the clumps were
also allowed to merge.
However, when employing sink particles,
gravitational interactions and also disc-protoplanet
interactions are altered as the clumps are not deformable.
Rather they are rigid bodies, 
with their interiors
being unresolved, a simplification that
completely suppresses the effect of shear and tidal forces.
This would indirectly affect the migration rate, as
tidal downsizing of clumps
\citep{boley-10, nayakshin-10} would slow migration considerably,
an effect that
has been shown to be statistically important in population
synthesis models \citep{galvagni-14, mueller-18}.
By missing the effect of tides, sinks
also miss an important  component: the angular momentum exchange phenomenology.
Further the choice of softening of the 
sink particles' potentials adds more
free parameters to the simulation
and 
alters the strength of clump-clump interactions.
In \citet{boss-12}, inward and outward
migration is observed,
and a few protoplanets are also 
scattered out of the disc, potentially leading to free-floating planets.

In this work we carry out a self-consistent study of migration
of protoplanetary clumps in fragmenting discs, examining for the
first time migration in both hydrodynamical and MHD simulations.
We intentionally do not replace the clumps with sink particles. This allows
us to fully assess the effect of clump-clump interactions
in migration, and to fully account for the interplay between accretion, tidal mass loss and migration.

\section{Methods}
\subsection{Numerical methods}
We analyze the simulations described in \citet{deng-20, deng-21, kubli}.
These are fully three-dimensional, self-gravitating simulations
of a protoplanetary disc and further include the magnetic field. The simulations have a very high resolution, 
with
a particle mass of \SI{2e-6}{M_{jup}}
in the MHD runs and
\SI{2.4e-5}{M_{jup}}
in the HD run. 
The 
gravitationally bound clumps which have sizes
in the range of \SI{0.1}{} to \SI{0.9}{AU}
are resolved with $\approx 5000 -100000$ 
particles, 
hence their internal structure is 
at least partially resolved.

Both the clumps (protoplanets) and the magnetic field emerge
self-consistently meaning their properties (mass / orbit of the clumps, strength and alignment of the magnetic field) are
not free parameters in the simulation but they emerge through
integrating the physical equations.
The simulation solves the MHD equations with self-gravity.
The magnetic component includes 
Ohmic diffusivity with 
$\eta = c_sH/25$
where $c_s$ and $H$ are the sound speed
and the disc scale height at the 
radial position of the fluid element.
The energy equation contains a cooling term with
a cooling time
$\tau_c = \beta \Omega^{-1}$
where $\Omega^{-1}$ is the local orbital time scale:

\begin{equation}
\frac{\partial \rho}{\partial t} + \nabla(\rho \mathbf{v}) = 0
\end{equation}
\begin{equation}
\frac{\partial \mathbf{v}}{\partial t} + \mathbf{v} \cdot \mathbf{\nabla} \mathbf{v} = -\frac{1}{\rho}\nabla(P + \frac{\mathbf{B}^2}{8\pi})%
+ \frac{(\mathbf{B}\cdot\nabla)\mathbf{B}}{4\pi\rho}%
- \nabla\Phi
\end{equation}
\begin{equation}
\frac{\partial \mathbf{B}}{\partial t} = \nabla \times (\mathbf{v}\times \mathbf{B}) + \eta\nabla^2\mathbf{B}
\label{eq-ohm}
\end{equation}
\begin{equation}
\frac{\partial U}{\partial  t} + \nabla(U \mathbf{v}) = -P\nabla \mathbf{v} - \frac{U}{\tau_\text{c}}\,.
\end{equation}

To solve the equations, we use GIZMO \citep{gizmo1, gizmo2, gizmo3}.
The simulations assume an ideal monatomic gas with adiabatic
exponent $\gamma = 5/3$ and equation of state $P = (\gamma - 1) U$.
The cooling constant $\beta$ is varied during
the earlier phases of the simulation 
(see section \ref{ch-setup}).
The gravitational softening is adaptive with a lower limit of
$0.05\,\text{AU}$ \citep{price-07}.
The central star is represented as a single sink particle
starting with $1M_\text{sun}$. Particles falling into
the sink radius, which is chosen to be $5\,\text{AU}$ are removed
and their mass and momentum is added to the star.

\subsection{Initial setup and fragmenting stage}
\label{ch-setup}
\begin{figure}
    \includegraphics[width=\columnwidth]{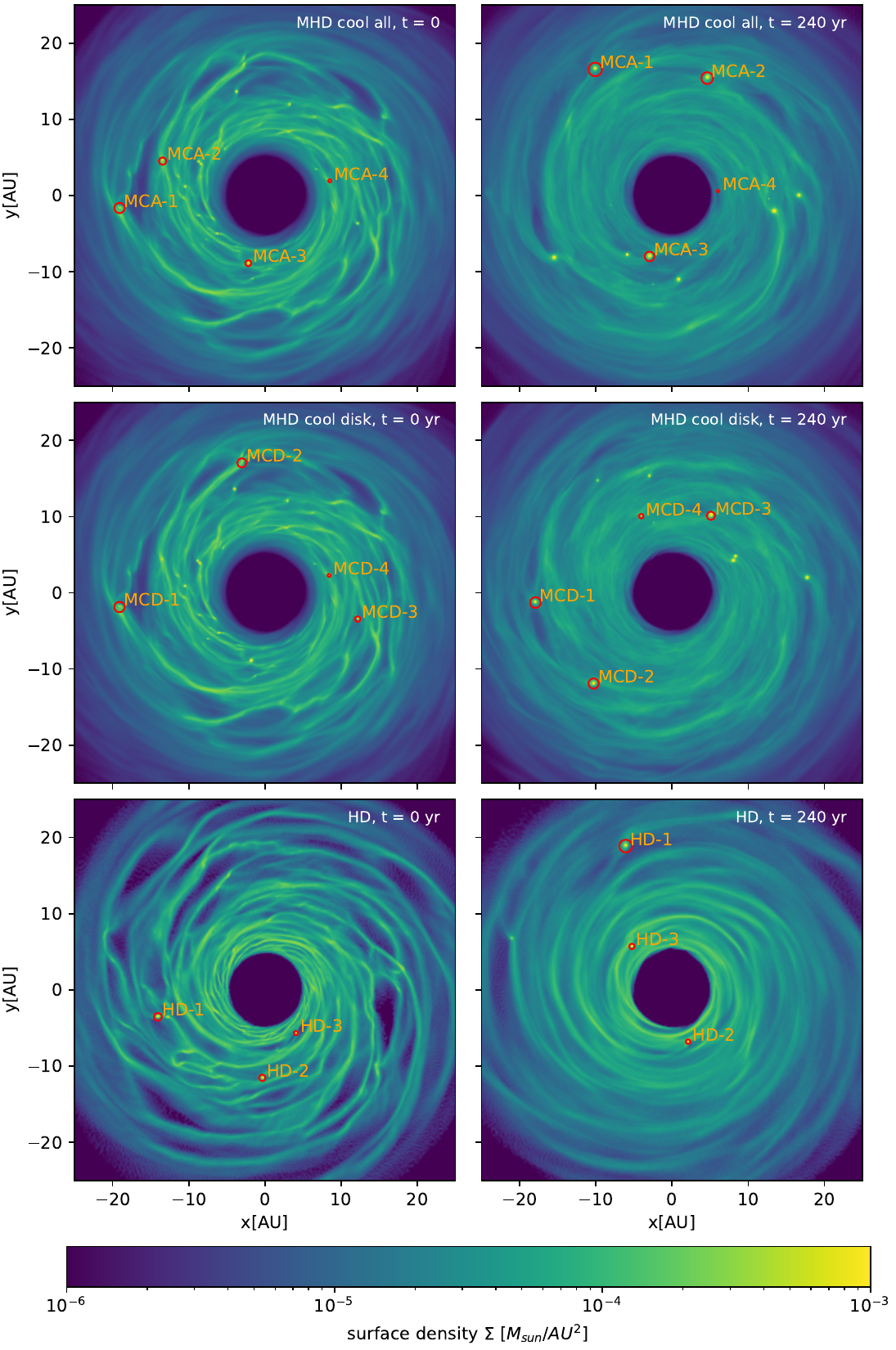}
    \caption{Surface density plots of the three runs at the beginning (left)
    and at a later stage $t = \SI{240}{yr}$.
    Top: \emph{MHD cool all}, including the magnetic field and applying 
    beta cooling on the whole system.
    Middle: \emph{MHD cool disc}, including the magnetic field and applying
    beta cooling \emph{only} to the disc, not the interior of the clumps.
    Bottom: \emph{HD}, neglecting the magnetic field and applying beta cooling to 
    the whole system.}
    \label{fig-density}
\end{figure}

The initialization stage is described in \citet{deng-20}.
The simulations start with an axisymmetric disc
with a radial extension of $5$--$25\,\text{AU}$, 
a surface density profile of $\Sigma \propto r^{-1}$
and a temperature profile of $T\propto r^{-1/2}$.
The local scale height of the disc was chosen such that
it is in pressure-gravity equilibrium.
The initial disc mass is
$\approx 0.1 M_\text{sun}$,
representing an earlier stage of a protoplanetary disc's 
evolution where gravitational instability is present.
A toroidal seed magnetic field was further introduced
which grows through the GI dynamo into an equilibrium configuration
over some orbits.
To avoid numerical fragmentation, 
a weak cooling rate of
$\beta = 8$
was used until the establishment of a spiral structure
through out the disc \citep{meru-bate-11, deng-17}.
The cooling was then increased to the 
desired
$\beta = 2\pi$ 
until the disc saturated to a quasi-equilibrium
state with vigorous MHD turbulence
powered by the spiral density waves
\citep{riols-19, deng-20}.
To trigger fragmentation we switched to a
faster cooling of $\beta = 3$
\citep{gammie-01, deng-17}.
This led to the formation of clumps over a range
of different sizes. 
This is described in \citet{kubli}
and \citet{deng-21}.
Most importantly, the fragments in the MHD case are of significantly lower mass than in the HD case, with many initial masses
below the mass of Neptune.
To follow the clump's evolution
and their interactions
the system was then evolved by switching
back to $\beta = 2\pi$
to avoid runaway fragmentation.
This corresponds to the time $t = 0$
in the surface density plots of fig. \ref{fig-density}.
As shown in \citet{kubli}, the clumps have a complex internal
structure, exhibiting flattened non-axisymmetric shapes and 
possessing significant angular momentum. Additionally, clumps
in MHD and HD simulations also show structural differences.
This suggests that treating them as sink
particles would not capture their mutual dynamical interaction
correctly, and also affect the exchange of angular momentum
with the disc. Since we want to understand the physics of migration
at some depth, we prefer to give up running for long timescales,
an inevitable issue as clumps contract and the integration time-step
becomes increasingly small, and rather make sure to fully capture the
dynamical interaction with the disc and between clumps for
as many orbits as we can follow the system.

\begin{figure}
    \includegraphics[width=\columnwidth]{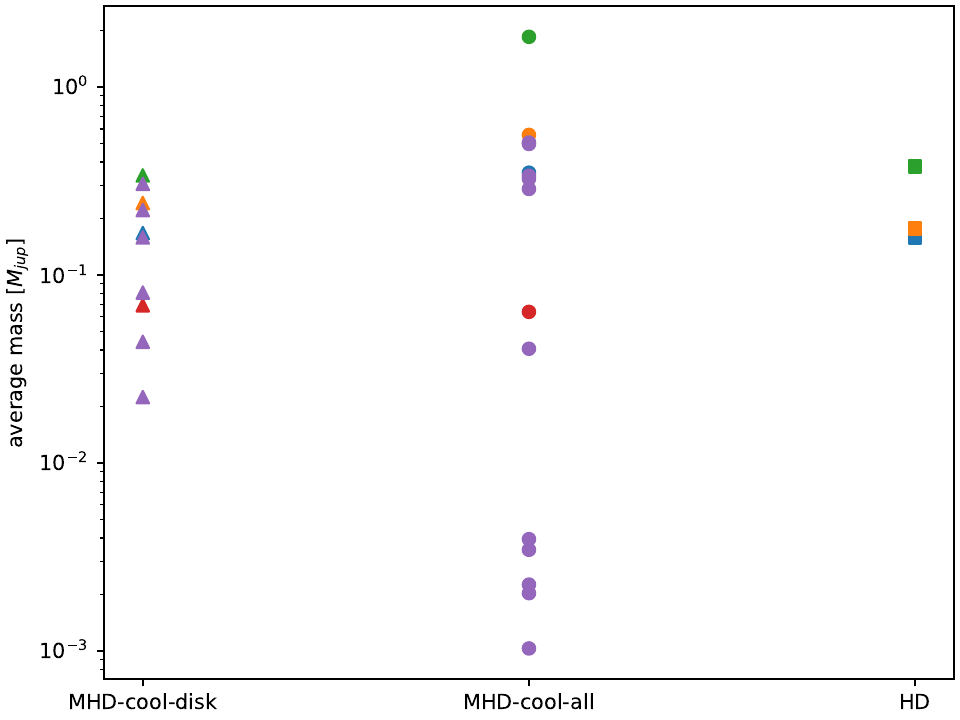}
    \caption{Average masses of the clumps over
    their lifetime in different runs.
    In the magnetized runs, 
    there are more fragments
    than in the purely hydrodynamical (HD) runs.
    The magnetic field also allows for fragmentation at lower masses.
    In \emph{MHD cool all}, 
    where the interiors of the clumps are also cooled, some clumps grow to much higher masses during the course of the simulation.
    \label{fig-clump-masses}}
\end{figure}
Due to the expected self-shielding in 
the interiors of the clumps 
it is physically reasonable to deactivate
cooling in the cores.
In a subset of our simulations this is  done by setting a density threshold
of $10^{-9} g/cm^3$ (see \citet{deng-21})
above which beta-cooling
is not applied.
The corresponding run is dubbed \emph{MHD cool disc (MCD)}.
Two companion runs are also used for the analysis.
In \emph{MHD cool all (MCA)} radiative cooling
remains active also in the clump interior. 
This leads to a more pronounced collapse
and a higher accretion rate on to the
clump making them more massive over time as shown in fig. \ref{fig-clump-masses}.
In \emph{HD} the magnetic field
is neglected hence this run takes
into account only the hydrodynamics
and the disc's self-gravity.
The surface density plots in fig.~\ref{fig-density}
show
a visible difference between the magnetized 
and the unmagnetized runs:
in the magnetized case the global spiral modes
are less pronounced but there is more 
small scale structure, including  higher frequency spiral modes \citep{deng-20}.

\subsection{Torque decomposition}
\label{ch-torque}
In this section we describe how we compute the torque that is exerted on the clumps and how we isolate its components.
For each clump, we start by computing the gravitational acceleration (specific force) $a_g$ on the clump's centre of mass.
Here we take into account all the disc's particles but exclude the other particles of which the clump itself consists.
The specific gravitational torque is then simply
\begin{equation}
    \mu = r \times a_\text{g}\,,
\end{equation}
with $r$ being the clump's position measured from the disc centre.
Assuming a circular orbit, the specific torque can directly
be related to a change in radius:
\begin{equation}
    (r \times a_\text{g})_z = \frac{\dot{r}}{2\sqrt{r}}\,.
\end{equation}

At each snapshot, the total torque is evaluated and the radius integrated.
Theoretically this should give us the exact evolution of the radial positions
since pressure or magnetic forces are not expected to be important on the clump's
orbital evolution.
However since the analysis is done on the snapshots that are available in discrete
time intervals 
of
$\approx 1.6\,\text{yr}$, some information is missing and the prediction will not be exact.
Further the clump's orbits are not purely circular but may have some eccentricity.
In practice the approximation works reasonably well for our purposes
(see section \ref{ch-results-prediction}).

In the  conventional linear migration theory giving
rise to Type I migration the contribution
to the torque outside the co-rotation region comes
from the sum of the resonant torques at the Inner and Outer Lindblad
Resonances (ILR and OLR) located, respectively, further in
and further out than the clump's orbit
\citep{LinPapaloizou1986}.
An additional contribution comes from material that 
co-rotates with the clump \citep{Tanaka2002}. 
Analogously,
in our analysis we dissect the torque on each clump 
into different components and
attribute the corresponding migration effects to each of them.
We define the following components:
\begin{itemize}
    \item The \textbf{clump-clump torque} 
    is the torque arising from interactions between 
    two or more clumps.
    \item The \textbf{disc torque} is the torque arising from the rest
    of the material (excluding other clumps). 
    We divide it further into different parts; the following components are all a part of the disc torque.
    \item The \textbf{outer disc torque} (part of the disc torque) 
    is the torque contribution
    from all matter on orbits further \emph{outside} than a Hill radius 
    as measured from the respective clump, while excluding
    clump-clump interactions.
    \item Vice-versa, the \textbf{inner disc torque} (part of the disc torque)
    is the torque contribution from all 
    matter on orbits further \emph{inside} than a Hill
    radius as measured from the respective clump, also
    excluding clump-clump interactions.
    \item Then we measure the remaining
    \textbf{torque from within
    1 Hill radius} (part of the disc torque) around the clump's orbit. 
    Also here, clump-clump interactions are excluded.
    We treat this part of the torque separately
    (not in the inner / outer torque) because
    the material on orbits close to the protoplanets
    is where we expect to be most of the contribution.
    Indeed, massive planets in massive discs can,
    in principle, enter the Type III or runaway regimes,
    in which the dominant contribution of the torques 
    comes from the material in the co-rotating region or near 
    to it (see e.g. \citet{papaloizou-07}).
    \item Although the above decomposition is complete
    (summing all contributions equals the total torque)
    we also show the \textbf{torque from within
    2 Hill radii} (part of the disc torque) around the clump's orbit (excluding clump-clump interactions).
    We do this to investigate what region around
    the clump's orbit is mostly responsible for the disc torque.
\end{itemize}

We want to stress here
that our simulations have a highly
non-linear flow as the disc is self-gravitating and
turbulent, due to gravito-turbulence (magnetized and HD runs)
and MHD turbulence (magnetized runs).
Both, gravitational stresses and Maxwell stresses are high in the MHD
simulations \citep{deng-20}.
Therefore, our flow conditions are at odds
with the assumptions made in linear theory.
However, the eventual contribution of resonant torques at ILRs and OLRs,
if relevant, is automatically included in, respectively, the inner and outer disc torque, which should be interpreted as a radially integrated
torque. It will still be instructive, a posteriori, to compare our
results with the expectation of linear torque theory, in particular
to compare with Type I migration torques and their timescales, and
to check if there is evidence of an important contribution at the
expected location of the resonances.

\section{Results}
\label{ch-results}
\subsection{Radial migration and torque-based predictions}
\label{ch-results-prediction}
\begin{figure}
\begin{subfigure}{\columnwidth}
\includegraphics[width=\columnwidth]{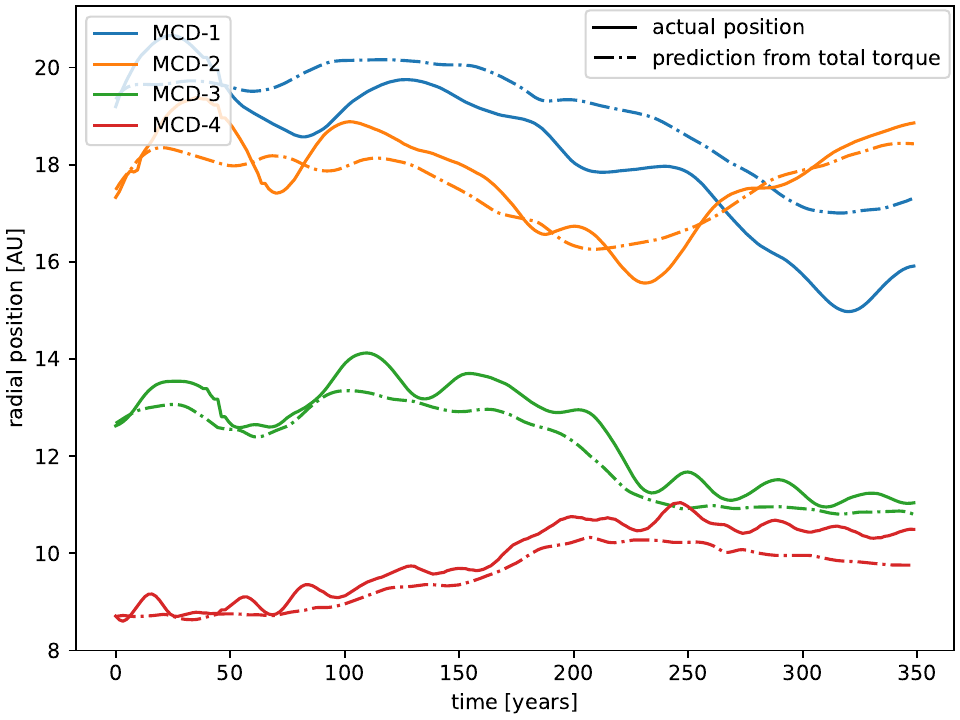}
\end{subfigure}
\begin{subfigure}{\columnwidth}
    \includegraphics[width=\columnwidth]{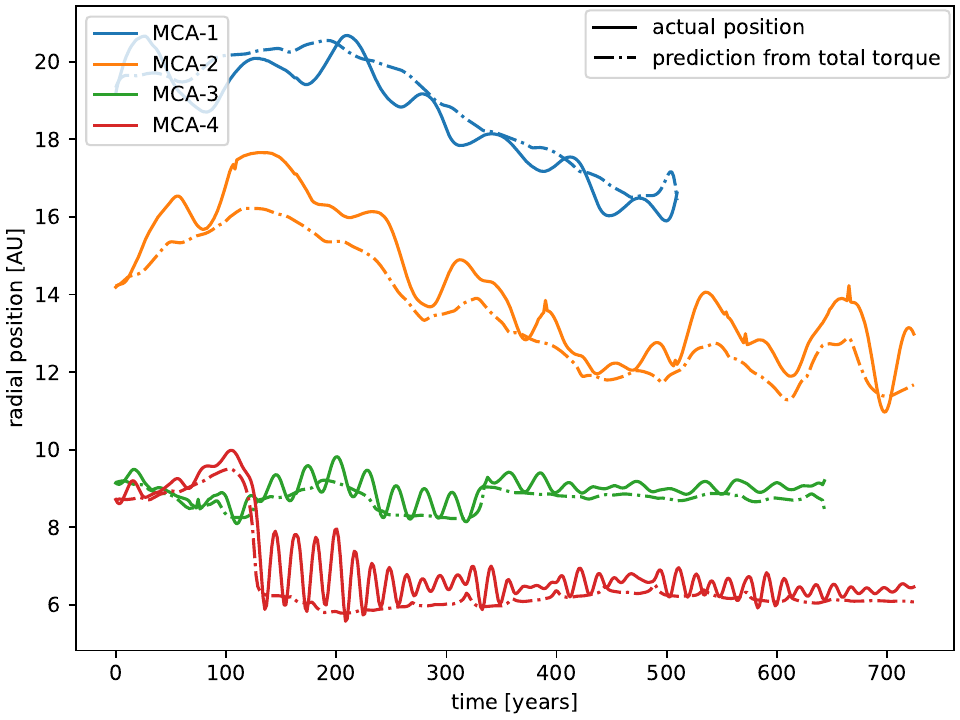}
\end{subfigure}
\caption{Radial positions of four exemplary clumps (solid lines) and 
their projected positions using the torque computed
from the simulation snapshots.
The evolution of the radial
position is well described 
using this method.}
\label{fig-predictions}
\end{figure}

We will start by assessing the robustness of our torque
measurement, and then consider separately the different torque components
defined in the previous section to uncover the in-depth
nature of migration for disc instability protoplanets.
For each different run we chose at most four exemplary clumps
which we show individually in the plots except
when we use the full sample for generating statistical
diagnostics, and use the 
same colours for them.
The solid lines in fig.~\ref{fig-predictions} show 
the observed radial positions of the clumps
in the simulation.
One can see that there is considerable change of the radial positions
over time.
For example, over \qty{\sim 5}{} orbits, clump MCD-1 
changes its radial position from \qty{20}{AU}
to \qty{16}{AU} and clump MCA-4 changes its position
rapidly from \qty{\sim 10}{AU} to \qty{7}{AU}.
Some clumps have a somewhat high eccentricity
(e.g. MCA-3 and MCA-4).
Besides inward-directed migration there is also 
outward-directed migration (clump MCD-4 of the ones shown).
The motions seem to have a large stochastic component.

First we verify that our method to estimate the torque at the (radial) location of the clump, described in section \ref{ch-torque},
is quantitatively consistent with the time evolution of the
semi-major axis of clumps observed in the simulation.
We show the result in fig. \ref{fig-predictions} 
for four exemplary plots in both MHD runs.
The dashed lines, representing the predicted positions of the clumps, closely follow the 
solid lines, representing the actual positions of the clumps.
We did not take into account the eccentricity of the clumps when integrating
the torque, but the prediction represents reasonably well an orbital average of
the clumps.
This already shows that the torque is mediated through
the self-gravity of the gas and we can neglect any
direct effects of the pressure forces and the magnetic field
on the migration of the clumps. 
Of course these forces are important when shaping the global
structure of the disc and therefore influence the final
torque arising from the self-gravity.

\subsection{The components of the torque}
\begin{figure}
\centering
\begin{subfigure}{\columnwidth}
	\includegraphics[width=\textwidth]{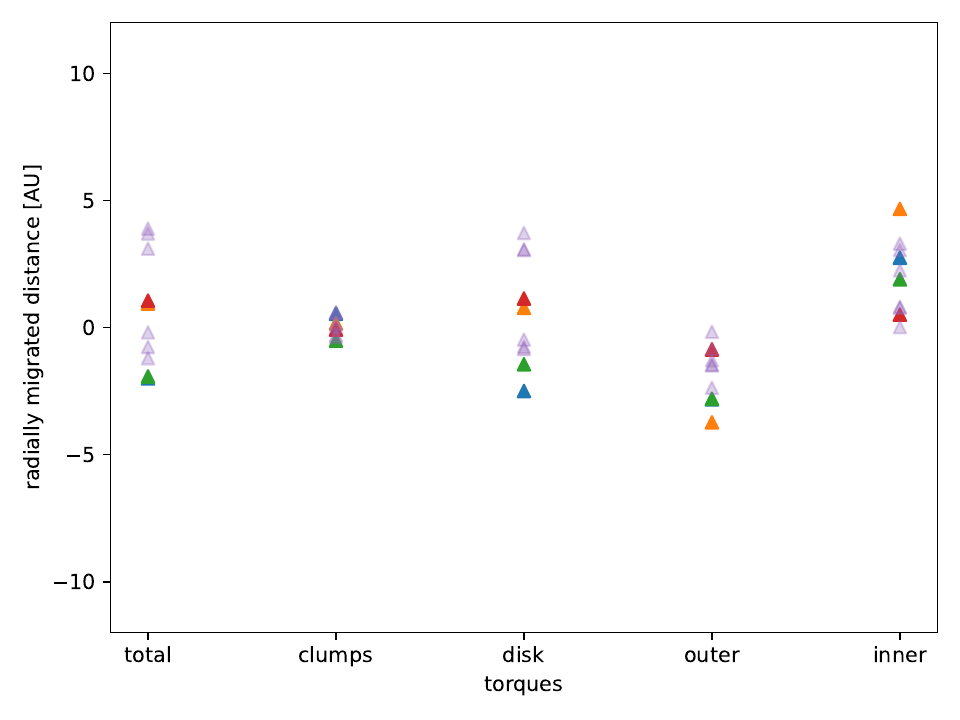}
 \caption{MHD cool disc}
 \label{fig-torque-compare-m}
\end{subfigure}

\begin{subfigure}{\columnwidth}
	\includegraphics[width=\textwidth]{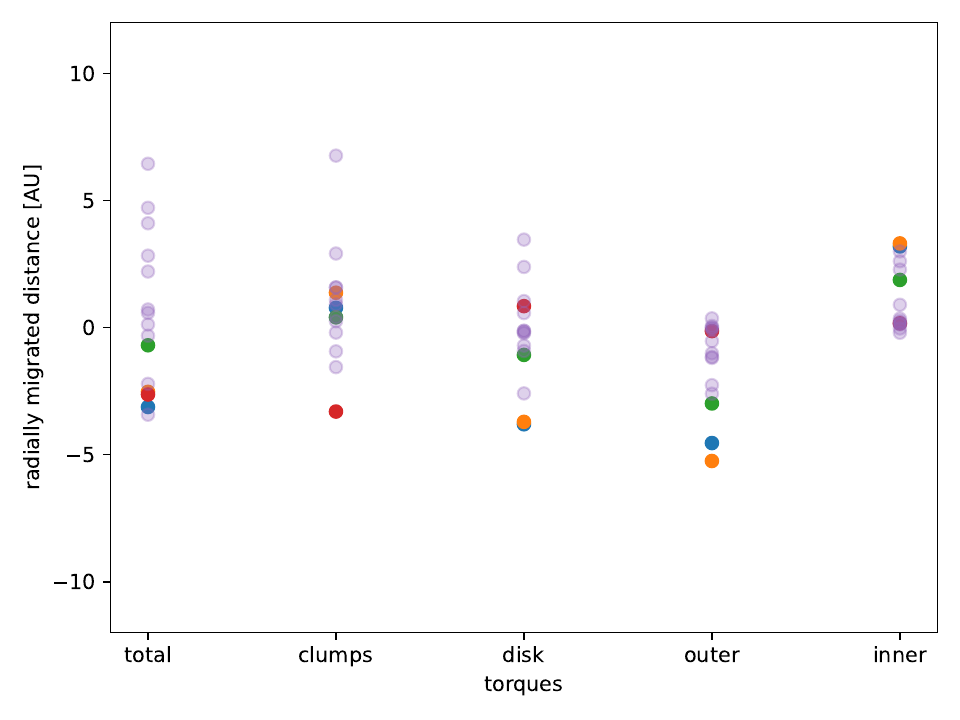}
 \caption{MHD cool all}
\label{fig-torque-compare-c}
\end{subfigure}
\caption{Change in radial position arising from different
components of the torque.
The effect of clump-clump interactions is 
much higher in the \mbox{\emph{MHD cool all}} case, consistent with the higher masses.
It can be seen that the outer torque leads to inward migration while the inner torque leads to outward migration in both cases.}
\label{fig-torque-compare}
\end{figure}

The effects of the different torque components
are summarized in fig.~\ref{fig-torque-compare}.
In both plots we show the radially migrated distance 
of each clump.
In the later plots of individual clumps they are marked
with the same colours.
All the other clumps found in the simulation, not shown individually in other plots, are shown in transparent purple.
The effects attributed to the different torque components
are shown in each column. 

The first column shows the total migrated distance
over the simulation time for each clump.
In all runs we have in- and outward migration. 
The next column represents migration due to 
clump-clump interactions. 
This is calculated by integrating the clump-clump torque 
(see section~\ref{ch-torque}).
The importance of this component is different between
the two magnetized runs \emph{MHD cool disc} and \emph{MHD cool all}.
In the run \emph{MHD cool disc} clump-clump interactions 
only account
for a fraction of the total migration whereas
in \emph{MHD cool all} clump-clump interactions play a significant role.
This is probably
due to the higher masses arising in \emph{MHD cool all} (see section~\ref{ch-setup}).
We note here that the run \emph{MHD cool disc} is probably more realistic
because cooling was turned off in the 
clump cores, which accounts for the self-shielding.
In such a case we expect migration
to be dominated mostly by the interactions with the
gas disc whereas clump-clump interactions 
add stochastic perturbations.

The other three columns to the right display the effect
of the disc. 
The outer and inner torque result from the previously 
defined decomposition of the disc torque 
(see section~\ref{ch-torque}).
The picture here is similar in both the magnetized and
the unmagnetized runs.
It can be noted that the outer disc torque is mostly
responsible for inward migration, thus it imparts a negative
torque component, whereas
the inner torque leads mostly to outward migration 
(see section~\ref{ch-disc-torque}).
This is consistent with expectations from Type I migration theory where 
the inner Lindblad resonance causes a positive torque and the 
outer Lindblad resonance causes a negative torque. However, we note
that a direct comparison with linear torque theory would not 
be well-posed since the gas flow in gravitationally unstable
discs is turbulent and there is no well-defined background flow,
rendering any perturbative treatment meaningless.
In the following sections we discuss the effect of the different
components in more detail.

\subsection{Clump-clump interactions}

\begin{figure}
\begin{subfigure}{\columnwidth}
    \includegraphics[width=\columnwidth]{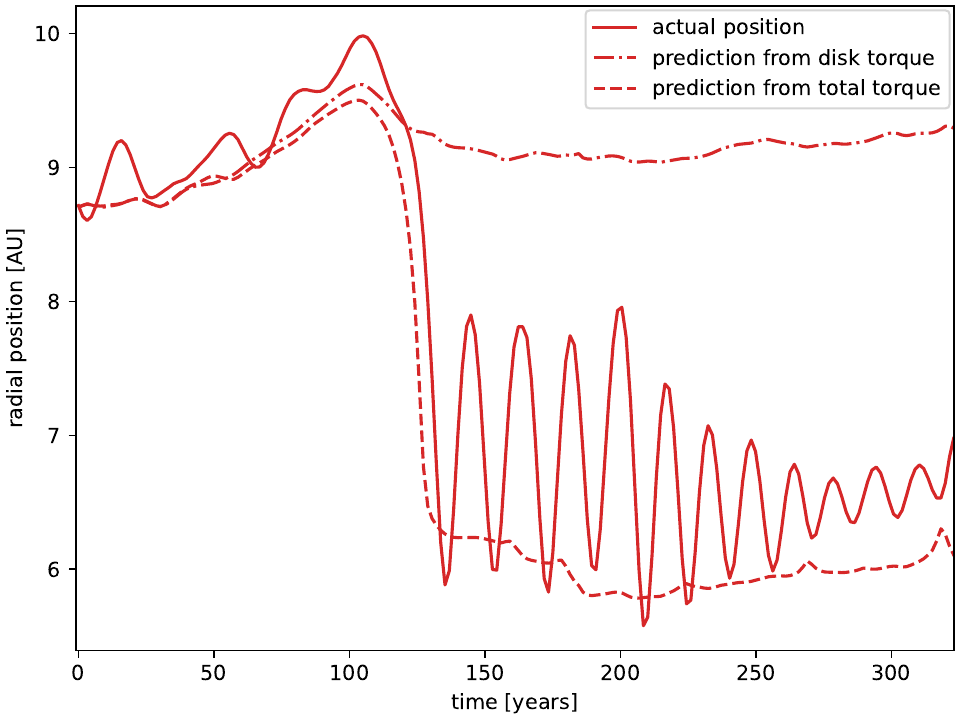}
\end{subfigure}

\caption{Evolution of the radial position
of clump MCA-4 during the interaction with MCA-3. 
The clump (solid line) is pulled inward during the interaction
and continues on an eccentric orbit. 
The position can only be correctly predicted by taking into account 
the clump-clump interactions, implying that the rapid change in 
radial position is caused by the latter.}
\label{fig-clump-clump}
\end{figure}

\begin{figure}
\includegraphics[width=0.7\columnwidth]{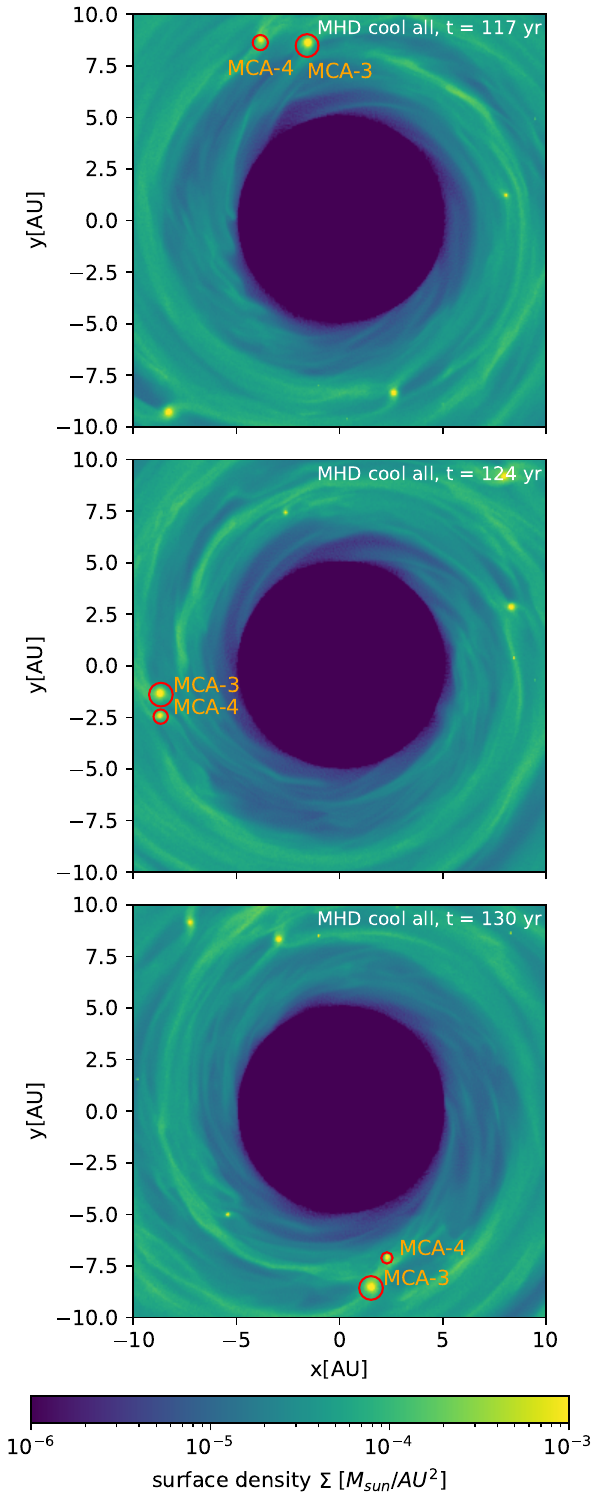}
\caption{Surface density plots of the disc during the interaction of 
clump MCA-4 with MCA-3. The more massive clump MCA-3 approaches the clump MCA-4 from the back
which is deflected inward and continues on an eccentric orbit.}
\label{fig-clump-clump-image}
\end{figure}

A potentially important component of the torque which is 
usually not present in conventional models in which
migration is treated as a pure planet-disc interaction
are clump-clump interactions 
(see also 
\citet{boss-12, boss-23}).
An example of such an interaction is shown in fig. \ref{fig-clump-clump}.
It shows the radial position of clump MCA-4 over time. 
The solid line shows the radial position of the clump in the simulation.
At a time of \SI{\sim 120}{yr}, the clump suddenly moves inward 
on an orbital timescale.
The orbital time of the clump is reduced from \SI{\sim 25}{yr}
to \SI{\sim 15}{yr}.
The two other lines in fig. \ref{fig-clump-clump} show the position
predicted from the torque using the procedure outlined in section
\ref{ch-torque}.
The dashed line which shows the predicted position taking into
account the total torque closely follows the actual position of the clump.
It is not exact which is likely due to the fact we only use the simulation
data in snapshots with discrete spaces of $10 / 2\pi\,\text{yr}$ 
and that we assumed zero eccentricity.
The plot also shows the dash-dotted line which is the predicted position
from the disc torque, meaning all torque components except for 
clump-clump interactions.
It can be seen that from the point on where the clump begins to move inward,
the disc torque predicts an almost constant radial position of the clump.
This shows that this rapid inward migration is really due to
clump-clump interactions.

The situation is shown in the density plots in fig.
\ref{fig-clump-clump-image} showing the inner part of the disc
at successive times.
The clump MCA-4 which is migrating inward and another close-by clump MCA-3 
are marked with the red circles.
It can be seen that the clump MCA-3 (being on a faster orbit) approaches MCA-4 
and thereby scatters it inward. The orbit of MCA-3 is less influenced as
it is a more massive clump.

\begin{figure*}
    \includegraphics[width=\textwidth]{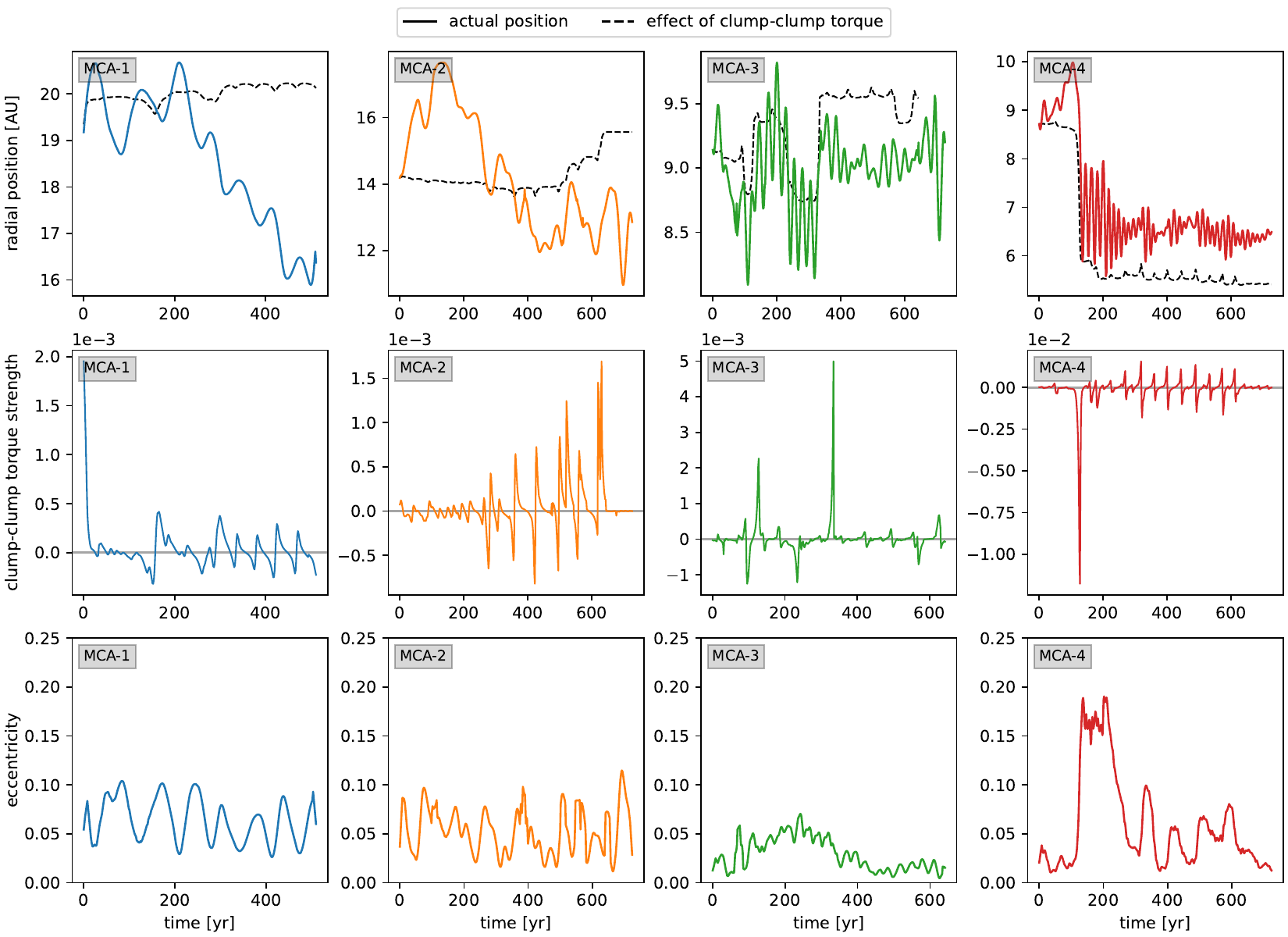}
    \caption{The clump-clump torque for the
    \mbox{\emph{MHD cool all}} case.
    Top: radial position of the clumps (coloured solid lines) and predicted positions taking into account only the effect of the 
    clump-clump torque (black dashed line).
    Middle: torque strength arising from clump-clump interaction for four different clumps.
    Bottom: eccentricity evolutions of the clumps.
    The clump-clump torque consists of short spikes leading to fast changes in the radial positions
    compatible with the short time of close-by encounters.
    Except for clump MCA-4 the clump torque alone does not seem to be the main
    driver of migration.
    Further, clump-clump interactions do not seem to lead to highly 
    eccentric orbits on the long-term.}
    \label{fig-clump-clump-torque}
\end{figure*}

Besides the fast inward migration, gravitational scattering also leads to a more 
eccentric orbit as can be seen in the oscillating behaviour of the radial
position in fig. \ref{fig-clump-clump}.
The eccentricity of the orbit is plotted in fig. 
\ref{fig-clump-clump-torque} (bottom right).
This is measured by evaluating the specific energy $\epsilon$ and the specific angular momentum $l$
of the clump and then using
\begin{equation}
e = \sqrt{1+2\epsilon l^2/(M_\text{star}^2G^2)}\,.
\end{equation}

It can be seen that around the time of interaction, the eccentricity
increases significantly.
However during a time of \SI{100}{yr} the orbit circularizes again.
This can be due to the interactions with the gas disc, or subsequent 
weaker interactions with other clumps
but the fact that 
clump-clump interactions
don't seem to increase the 
eccentricity over the long term (bottom plots in fig. \ref{fig-clump-clump-torque}), suggests that the circularization is caused
mainly by the interaction with the disc.
This result suggests that although clump-clump interactions may lead to an increase
in eccentricity of the clumps, one may still expect on average 
low-eccentricity
orbits ($e=0.1-0.2$) as the eccentricity can quickly diminish again. We note that the values of the eccentricity found in these
simulations are similar between magnetized and non-magnetized discs, and
are in agreement with earlier results in the literature (e.g. 
\citet{mayer-2002, mayer-2004}).

Fig. \ref{fig-clump-clump-torque} shows the action of the clump-clump torque.
For four clumps it shows the torque strength over time (top) and the radial
position of the clump (solid line, bottom) together with the predicted position of
the clump taking into account only the clump-clump torque (dashed line, bottom).
It can be seen in the top part of the figure that the clump-clump torque acts on a short timescale and often 
switches the sign periodically. 
This is due to the clump overtaking / being overtaken by another clump
on a close orbit
and thus the direction of the gravitational force exerted from the other clump
switches.
In the bottom of fig. \ref{fig-clump-clump-torque} it can be seen
that this leads to fast, almost discrete changes in the predicted radial position
of the clump.
The magnitude of the clump-clump torque and thus its importance relative to the other 
components of the torque differs much between the clumps:
in the case of clump MCA-4 the effect of the clump-clump interactions dominate over other
components of the torque; in \mbox{MCA-2} and \mbox{MCA-3} both are important and in \mbox{MCA-1} clump-clump interactions
are only of minor importance.

\begin{figure}
\begin{subfigure}{\columnwidth}
    \includegraphics[width=\columnwidth]{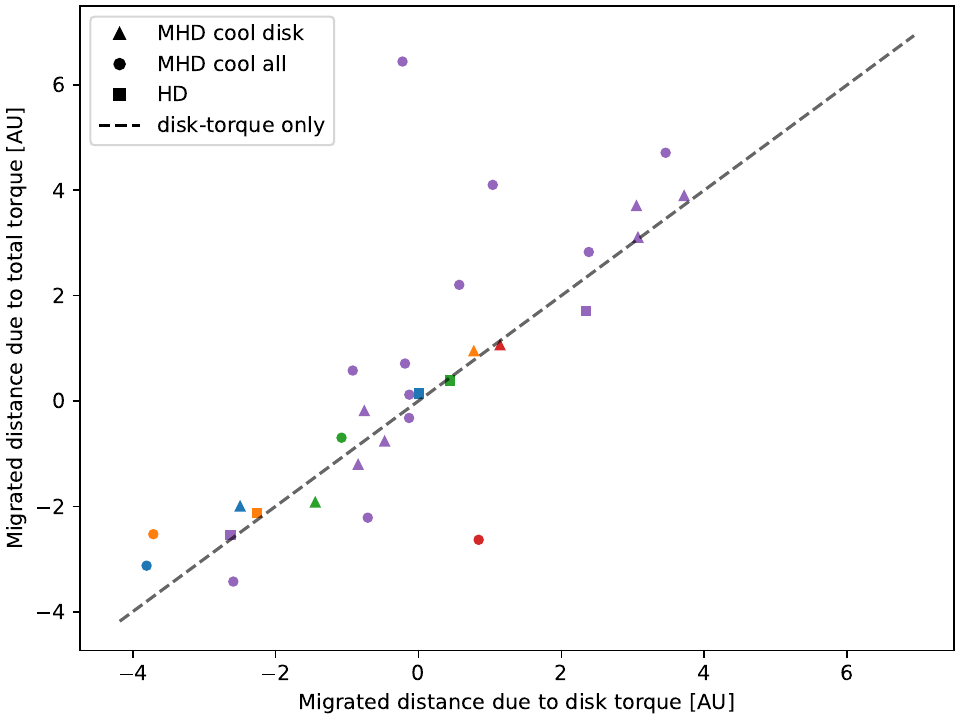}
\end{subfigure}

\caption{Migrated distance due to the total torque vs 
migrated distance due to the disc torque for each clump
of the three runs. Deviations from the diagonal line signalize
situations in which clump-clump interactions can not be
neglected in 
explaining the clump's migration.
For most clumps the disc torque is more important,
however
for some clumps especially in the 
\mbox{\emph{MHD cool all}}
run, the 
clump-clump interactions dominate, as shown by the larger
deviations of the corresponding data points with respect to
the diagonal line.}
\label{fig-disc-torque-expl}
\end{figure}

Fig. \ref{fig-disc-torque-expl} quantifies the overall importance of the clump-clump torque taking
into account the different runs.
It compares the migrated distance of each clump over the simulation to the distance
it would have migrated if the disc torque alone would have interacted with it.
The clumps of different runs are marked with different symbols, the clumps
in blue, orange, green and red are the ones we show in the grid plots for each run.
The dashed line is the identity, clumps on this line did not experience an overall
change in their radial position due to clump-clump interactions.
This is also the region where most of the clumps are in the plot.
However especially in the \emph{MHD cool-all} case there are some clumps that are far
away from the line and for which the disc torque yields a secondary
contribution.
Deviations from the line occur both from above and from below, signaling that clump-clump interactions can lead to  both a positive and a negative torque, and thus contribute to both inward and outward migration.
Overall, we infer that clump-clump interactions add a degree of randomness to the dynamics of the system which, in some instances, and
for a limited time span, can affect significantly the dynamics of the clumps.

Nevertheless, in the \emph{MHD cool disc} and the 
\emph{HD} case however, the disc-driven torque is the most important component. This is presumably because both runs have a lower number of clumps compared to the \emph{MHD cool all} simulations, thus reducing the occurrence of clump-clump interactions and also because of the 
higher masses in the \emph{MHD cool all} run. In the next section we analyze and discuss the role of 
the disc-driven torque in migration at greater depth.

\subsection{Decomposition of the disc torque}
\label{ch-disc-torque}
\begin{figure*}
    \includegraphics[width=\textwidth]{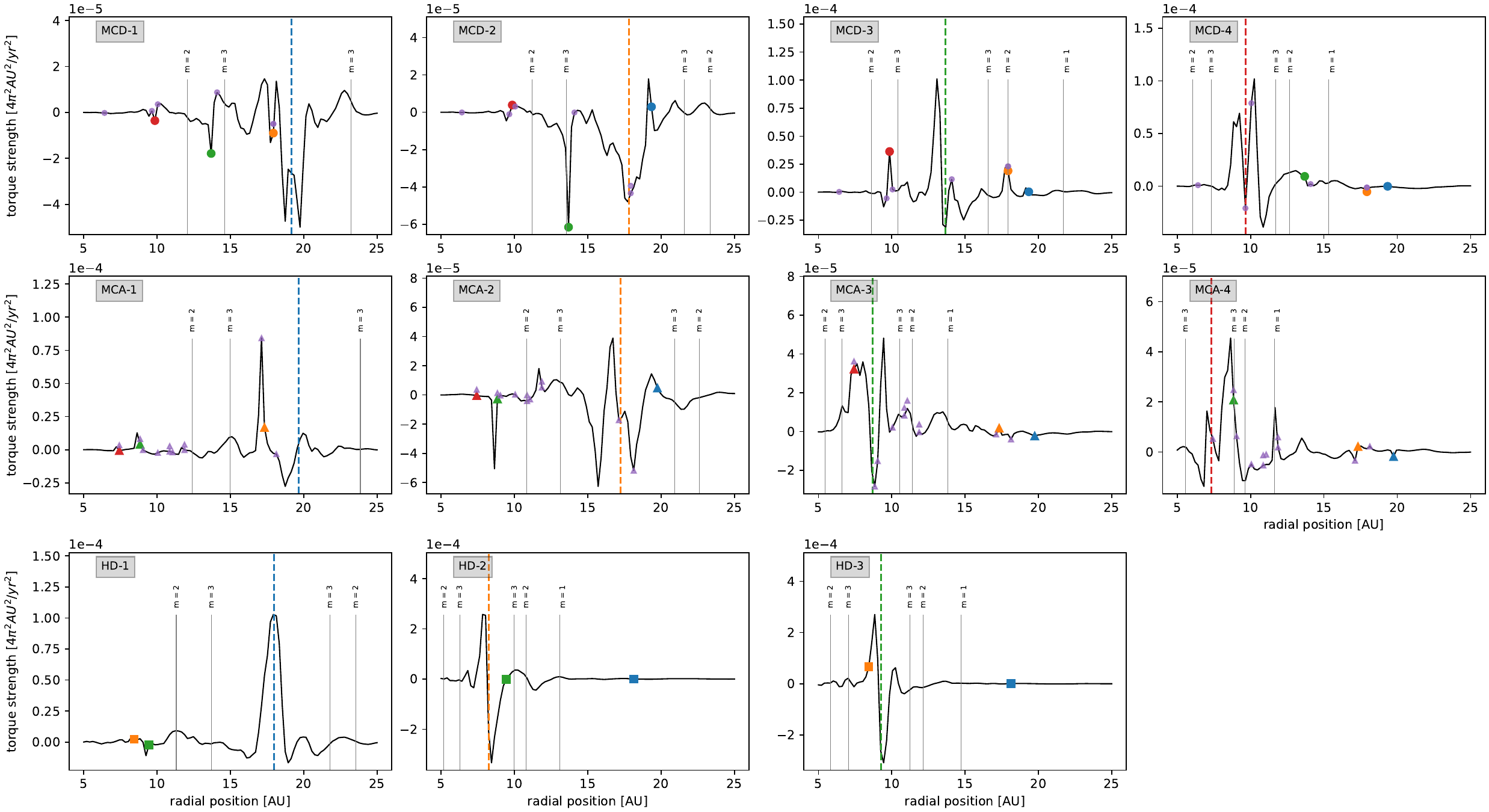}
    \caption{Profiles of the torque acting on each clump at time 
    $t \sim \SI{160}{yr}$.
    The y-axis shows the torque strength and the 
    x-axis shows the radial position in the disc, revealing
    where the torque is coming from.
    The position of the respective clump is shown 
    with the dashed lines and the radial positions of the other clumps
    are shown with the coloured dots.
    The position of the lowest Lindblad
    resonances are shown with the labelled,
    thin grey lines.}
    \label{fig-torque-profiles}
\end{figure*}

\begin{figure*}
\includegraphics[width=\textwidth]{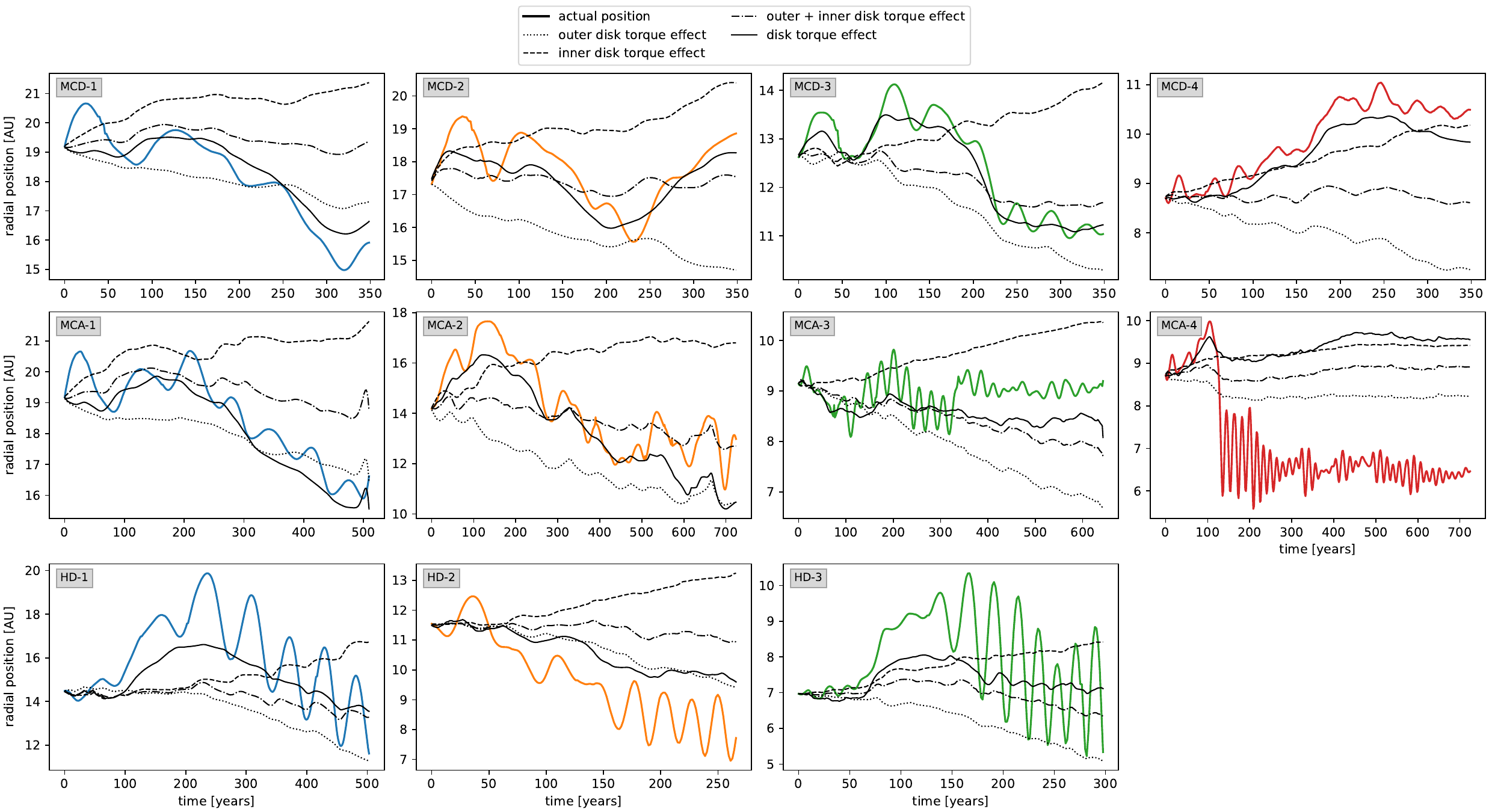}
    \caption{Comparison of the effects of the inner
    and the outer disc torque for different clumps. The evolution of the actual
    radial position of the clumps is represented by the coloured solid line. The outer disc torque (dotted line) tends
    to pull the clumps inward whereas the inner 
    disc torque (dashed line) pulls them outward.
    Together (dash-dotted line) these two torques are
    for some clumps in alignment with
    the total effect of the disc torque
    (black solid line),
    however
    the torque from $2 r_\text{hill}$ 
    shown in fig. \ref{fig-disc-coorb-compare} %
    resembles the evolution of the clump's radial position much more closely.
    }
    \label{fig-inner-outer-compare}
\end{figure*}

\begin{figure*}
    \includegraphics[width=\textwidth]{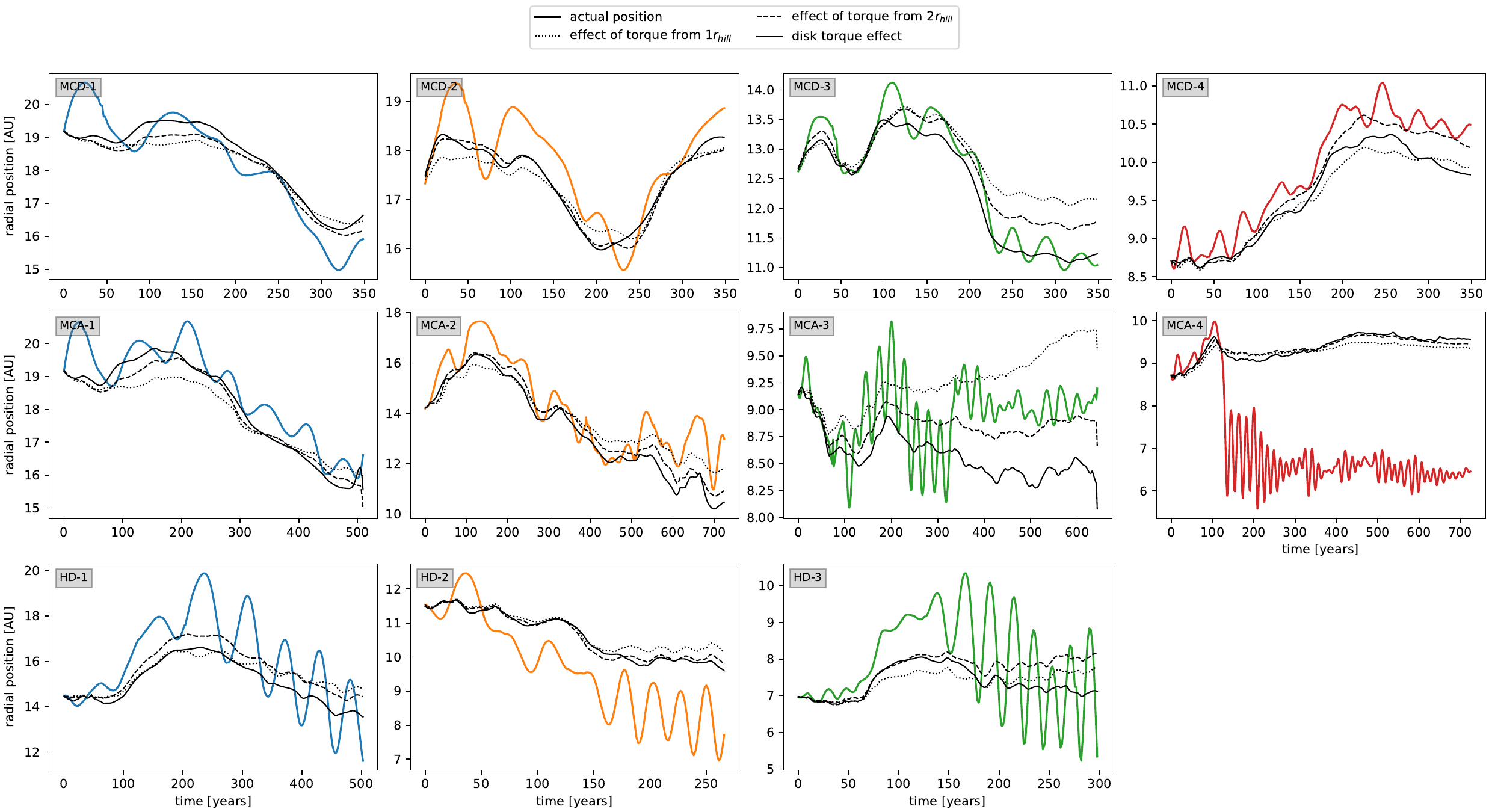 }
    \caption{
    Comparison of the effects
    of the torque from up to
    $1 r_\text{hill}$ (dotted line)
    around the orbit of each clump and
    the respective torque from up to $2 r_\text{hill}$ (dashed line; contains the former)
    to the total disc torque 
    representing the total torque without clump-clump interactions (dashed line, contains the former two).
    The evolution of the actual
    radial position of the clumps is represented by the solid line. 
    The torque from $2 r_\text{hill}$
    seems to account for most of the migration when neglecting
    clump-clump interactions
    and the other components of the disc torque (outer and inner disc torque) are only occasionally important.}
    \label{fig-disc-coorb-compare}
\end{figure*}

In this section we attempt to determine to what extent the orbital
evolution of the clumps, and thus the time-integrated effect of the
torque, can be or can not be reproduced by considering only a subset
of the disc torque components, and neglecting the episodic effect
of clump-clump interactions.
We should note that, in principle, even when clump-clump encounters
are negligible, multiple clumps are always present in the disc and 
might perturb the orbit of a given target clump because of their
long-range gravitational effect. We quantify the role of such ``far-field'' perturbations by analyzing the radial profile of the torque
acting on several target clumps. We perform the analysis for different clumps and at different times, but show the results for only one representative time in this paper, in fig. \ref{fig-torque-profiles}, since we found the result to be similar in the various cases considered.
We compute the profiles by dividing the disc
into $100$ radial bins.
For each fluid element in the system we then calculate in which
bin it lies and add its contribution to the torque to the 
corresponding bin.
Fig. \ref{fig-torque-profiles} shows that usually the dominant torque component originates from disc material relatively close
to the target clump, which does not correlate with the location
of other clumps. The latter, especially in the magnetized runs (see the panels for ``MCA-'' clumps in the figure) clearly produce both positive
and negative torque components, as there are bumps correlated with their
locations. The amplitude of these bumps is  usually much lower than that of the dominant torque component in the co-orbital region around
the target clump. When torque peaks appear near clumps, they are transients, as seen by comparing 
fig. \ref{fig-torque-profiles}
to fig. \ref{fig-torque-profiles125} in two adjacent 
time-steps (see e.g. clump MCA-1). Instead, the high amplitude torque, negative or positive, in the co-orbital region, is a robust feature, both in the HD and 
in the MHD runs. Additionally, we do not notice systematic differences in the torque profiles between MHD and HD runs, rather the specific
dynamical state of the system, especially the perturbations induced
by other clumps, is what determines, at any given time, the shape
of the torque profiles.

In fig. \ref{fig-torque-profiles}
we also show the location of the low-order Lindblad
resonances, which, in standard Type I migration, are the locations
where the exchange of angular momentum with the disc is maximal and
the torque is stronger. Clearly there are no signifiant features
in the torque profiles associated with these resonance in our
simulations. The rare instances in which a prominent torque feature
is seen at a similar location are mostly associated with a perturbing clump.
We also note that the dominant torque does
not come from material at Lindblad resonances, thereby highlighting
how the nature of disc torques in  self-gravitating, fragmenting 
discs is different from classical Type I migration in 
low-mass,
non-self-gravitating discs. 
The local nature
of the torque in self-gravitating discs was pointed out also by
\citet{malik-15},
who studied individual migrating protoplanets
in marginally unstable but smooth (non-fragmenting) discs using a very different hydrodynamical code, the finite-difference polar grid code
FARGO-ADSG \citep{masset-00, baruteau-fargo-08},
which solved the fluid equations in the co-rotating frame and
in two dimensions.

We now consider the effect of the different components of the disc torque over time, rather than at fixed time, and divide them
into three different components 
as described in section \ref{ch-torque}: inner disc torque, outer disc torque, and a torque from within one Hill radius 
around the respective clump's orbit. 
In order to compute these components, we first determine
for each fluid element if it is part
of another clump. 
In that case we don't count it here and it
only contributes to the clump-clump
torque.
We then determine in which region the fluid
element lies:
if its orbit is at a distance within 
a Hill radius around the clump's orbit
we count it towards the ``Hill torque''.
If the fluid element is on an 
orbit further than a Hill radius of the clump's orbit
it counts either to the outer
or the inner torque depending
on if it is inside or outside
the clump's orbit.
The contribution due to each
component is then determined by summing
the gravitational torques from each
fluid element belonging to that component.
We further compute the torque from
within $2$ Hill radii around the respective clump's orbit 
(see section \ref{ch-torque}).
We note that this component overlaps with the
outer / inner disc torques.

In fig. \ref{fig-inner-outer-compare}
we show the effects of the inner and outer disc torques
and compare them to the effect of the (total) disc torque
and to the actual migration of the clumps.
First it can be seen that the for most clumps
the outer disc torque (dotted lines) is negative and therefore
its isolated effect is an inward migration of the clump.
Conversely, the inner disc torque (dashed lines) is mostly positive
and therefore leads to outward migration.
This is analogous to the Type I picture where the OLR and ILR
cause torques in opposite directions.
In the plot we also show the sum of these two contributions
(dash-dotted lines). It can be compared to the effect of the (total) disc torque (black solid line).
Overall these parts of the torque are not in a good alignment
with the total disc torque (except for MCA-3)
hinting towards a greater importance of the material on
orbits close to the clump.

Therefore we now focus on the torque components near 
the clump's orbit.
Also the radial torque profiles in fig. \ref{fig-torque-profiles}
suggest that these parts are most often the dominant contributions;
apart from clump-clump interactions.
This may come from the gas flowing asymmetrically through the clump's orbit as well as by material in the circumplanetary disc 
assembling around the clumps \citep{szulagyi-17}.
For migrating planets in non-self-gravitating discs it has been
shown that fast, Type III migration takes place. This is driven
by the gas flowing through the horseshoe region, which yields
a negative torque, which can be compensated, or not, 
by a positive torque induced by material in the circumplanetary disc
\citep{papaloizou-03, papaloizou-07}. If the mass
flowing through the horseshoe region is larger than the sum
of the masses of the protoplanet and circumplanetary disc, then
this so-called ``mass deficit'' gives rise to a net negative torque
and the fastest mode of Type III migration, called runaway migration,
arises. For individual migrating protoplanets in massive self-gravitating discs, it has been shown that the conditions for 
runaway migration are naturally satisfied
\citep{malik-15}.
If, temporarily, the mass deficit becomes negative, then the positive
torque induced by the circumplanetary disc material can, in principle,
generate outward migration.

In our analysis we find that considering the contribution of the material within 2 Hill radii which
we defined as
\begin{equation}
    r_\text{hill} = r\sqrt[3]{m/m_\text{star}}\,,
\end{equation}
with $m$ the mass of the clump
and $r$ its radial position,
neatly identifies
the region mostly responsible for the disc torque. When comparing the 
torque induced by the material within twice the Hill sphere with the total disc torque, we find that the former  provides the most important contribution to the orbital evolution of the clumps, as shown in fig. \ref{fig-disc-coorb-compare} by the fact that it best reproduces the clumps' radial distance evolution overall.
We stress that this is true both for inward and outward migrating 
clumps.

\begin{figure}
    \centering
    \includegraphics[width=\columnwidth]{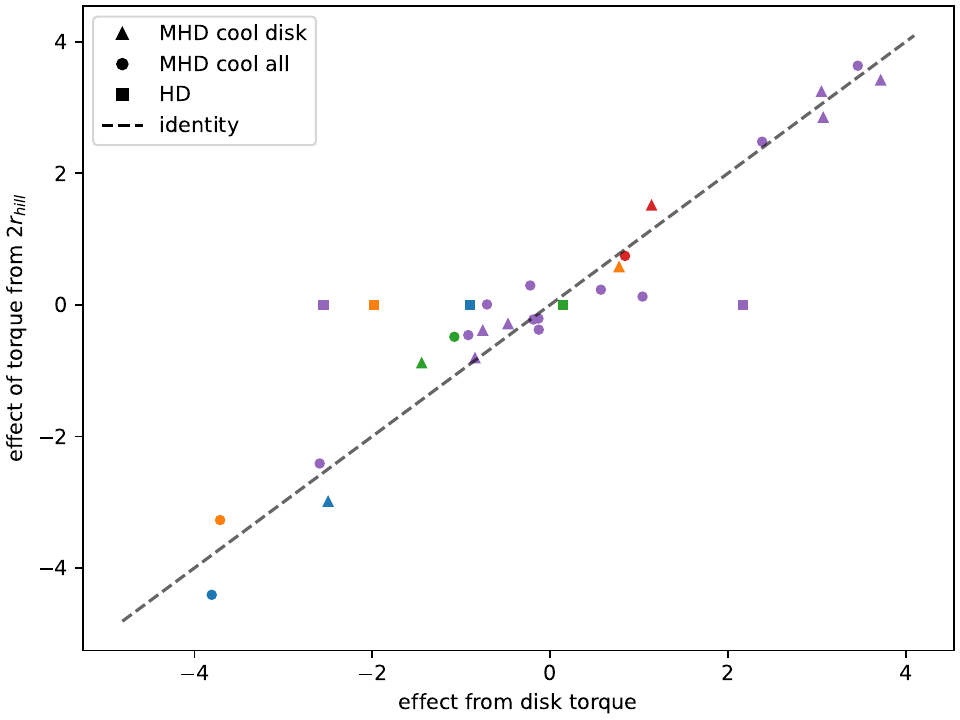}
    \caption{The y-axis shows the migrated distance due
    to the torque arising from the matter on orbits of 
    at most two hill radii apart from the clump's orbit;
    the x-axis shows the migrated distance due to the disc torque 
    for each clump.
    Most of the disc torque can be explained by the ``$2r_\text{hill}$'' torque, shown by the fact that most 
    clumps lie close to the diagonal line.}
    \label{fig-hill-torque-expl}
\end{figure}

Fig. \ref{fig-hill-torque-expl} shows that the total migrated distance of each
clump is very well reproduced by considering only this torque
component. 
For most of the clumps,
already the torque from within 1 Hill radius
around the clumps' orbits
reproduces the (total) disc torque contribution quite well.
The results are also in excellent agreement with the findings
of 3D SPH simulations of migrating black hole perturbers in
self-gravitating circumnuclear discs in galactic nuclei
\citep{mayer-13, souza-20}, suggesting that the local nature of the
torque is a general property of this regime. Note that in the latter
works, as in \citet{malik-15}, perturbers were evolving in a marginally
unstable but non-fragmenting disc,
hence torques were only generated
by the surrounding disc.
In our case, as we have extensively 
shown, clump-clump encounters and perturbations also play a role.

\begin{figure}
    \centering
    \includegraphics[width=\columnwidth]{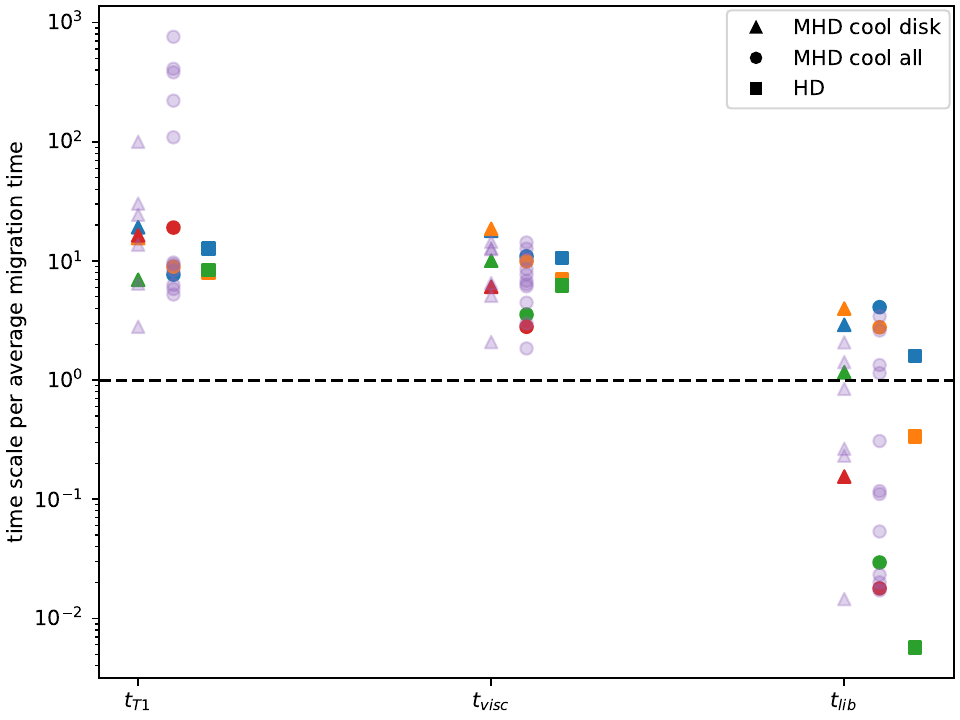}
    \caption{Comparison of different timescales
    relative to the measured migration time.
    For most clumps the Type I (Lindblad) migration timescale 
    $t_\text{T1}$ and the viscous timescale $t_\text{visc}$ (Type II migration) is much longer than the migration time.
    Only Type III migration, 
    is compatible with the observation as the libration timescale is similar or shorter than the migration timescale.}
    \label{fig-timescales-rel}
\end{figure}

Fig. \ref{fig-timescales-rel} shows different timescales relevant for
migration and compares them
to the actually measured migration time.
We first calculate for each clump the 
mean migration rate by measuring
$s=\left<\left|\dot{r}_\text{p}/r_\text{p}\right|\right>$ where $r$ is the clump's radial position.
The average is taken over each snapshot of the simulation.
We take the absolute value of
the term in the brackets because 
we have both in- and outward migration.
The migration time is then $t_\text{mig} = 1/s$.
We then use this time as a reference
for the other timescales.

We calculate the Type I 
Lindblad torque  from \citet{paardekooper-10}:
\begin{equation}
    \Gamma = \Gamma_0 / \gamma \left(-2.5 - 1.7\beta + 0.1\alpha\right) \,,
\end{equation}
with 
\begin{equation}
    \Gamma_0=\frac{m^2}{m_\text{star}^2 h(r_\text{p})^2} \Sigma(r_\text{p})r_\text{p}^4\Omega(r_\text{p})^2 \,.
\end{equation}
Here, $\alpha$ and $\beta$ 
are the negative power-law exponents of the surface density profile ($\Sigma \propto r^{-\alpha}$) and the temperature profile ($T \propto r^{-\beta}$),
$\gamma = 5/3$ is the adiabatic index, 
$m$ the mass of the clump,
$m_\text{star}$ the mass of
the central star,
$h$ the aspect ratio of the disc, $\Omega$ the angular velocity of the disc
and $r_\text{p}$ the radial position of the planet.
It can be seen that
for most clumps 
the Type I Lindblad migration timescale $t_\text{T1}$ is 
$\approx 10$ 
times more than $t_\text{mig}$, for some clumps
much more.
Further, the Lindblad torque
does not predict the correct
direction of migration.
This is already expected
from our results as we showed
that most of the contribution
comes from material on orbits
close to the clump.

We also estimate the viscous 
timescale by
\begin{equation}
    t_\text{visc} = \frac{r_\text{p}^2}{\alpha c_s h}\,,
\end{equation}
with $\alpha$ the viscous stress; in this case 
(since there is no viscosity in the simulation) an effective
value coming from gravitational, Reynolds and Maxwell stresses.
In our case
$\alpha \approx 0.2$ for the magnetized
runs and $\alpha \approx 0.1$ for 
the unmagnetized run \citep{deng-20}.
The variable $c_s$ is the sound speed of the disc.
It can also be seen that 
for most clumps  
$t_\text{visc}$ is $\approx 10$ times longer than the migration time,
thereby also 
ruling out Type II migration \citep{lin-86, cloutier-13}
through gap-opening.

We finally compute the libration
timescale \citep{baruteau-08}
\begin{equation}
    t_\text{lib} = \frac{8\pi r_\text{p}}{3\Omega r_\text{hs}}\,,
\end{equation}
with $r_\text{hs}\approx 2.5 r_\text{hill}$ \citep{paardekooper-09} the
width of the horseshoe.
It can be seen that for many clumps
$t_\text{lib}$ is of the 
same order as the migration time and for many clumps it is much shorter which means that the observed migration is consistent
with fast Type III migration similar to what was observed in 
\citet{malik-15}.

\begin{figure*}
\includegraphics[width=\textwidth]{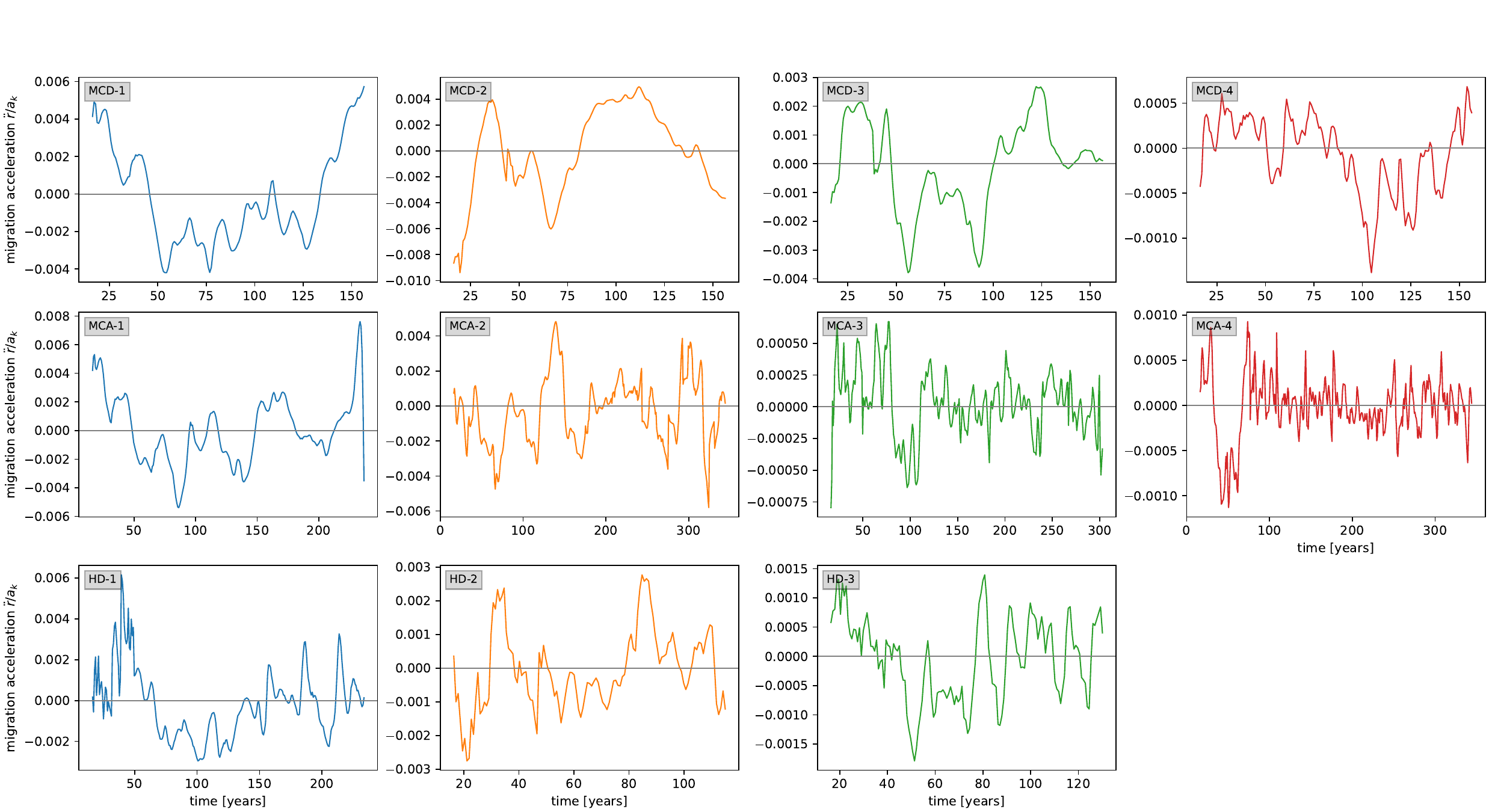}
\caption{Acceleration of 
the radial migration for a representative subset of the clumps.
The acceleration is normalized
to the Keplerian 
orbital acceleration and smoothed over
$\approx 60\, \text{yr}$ corresponding to 
an orbital time at $15\,\text{AU}$.
For most clumps, the sign of the acceleration
is alternating appreciably, in line with the stochastic migration
scenario.}
\label{fig-mig-acceleration}
\end{figure*}

We caution that, despite the apparent consistency with the Type III 
migration scenario, the  flow dynamics in fragmenting discs is not
only very different from the conditions in conventional simulations
for migrating planets in non-self-gravitating discs, 
but also
different from the conditions in self-gravitating but non-fragmenting
discs such as in \citet{malik-15}.
During fragmentation the gravitational
stress is maximal
\citep{lodato-04, deng-20},
hence the
departures from the axisymmetric Keplerian flow are the largest possible,
with significant spatially and time-varying local radial motions. As a result, one expects the migration rate to fluctuate.
We therefore neither expect monotonic runaway migration
(as often documented in the literature
for non-self-gravitating discs for the 
first $\approx 20$ orbits \citep{peplinski-doctoral}),
nor saturated steady state migration.
Finally, fig. \ref{fig-mig-acceleration}
shows the second derivative of the clumps' 
radial positions 
(acceleration of migration),
for representative cases.
In the case of runaway migration, one would observe
a monotonic and possibly exponential 
increase of acceleration; whereas in a steady state 
migration, the acceleration would vanish.
Instead, the second derivatives of our clumps
are fluctuating quickly, often changing sign on an orbital time.
In the next section we discuss this point further.

\section{Discussion and Conclusions}
\label{ch-discussion}

We investigated planetary migration in 
self-gravitating and also magnetized discs.
To this end, we analyzed 
high-resolution simulations in which 
protoplanetary clumps formed through fragmentation,
both in presence or absence of a magnetic field.
We intentionally refrained from employing sink particles for the clumps
to avoid spurious numerical effects.

For our analysis we computed the gravitational torque
exerted on the clumps in discrete snapshots and computed
a predicted radial position in each snapshot.
This method showed good agreement with the 
actual radial position of the clumps.
We then dissected the torque into multiple components
in order to associate the radial evolution of the
clumps with effects due to different components of the
surrounding flow. In particular, we separated the 
torque due to clump-clump collisions from that due to the
disc.

Our main results can be summarized as follows:
\begin{itemize}
\item Clump-clump interactions add stochastic perturbations
to the radial evolution of the clumps, but for most clumps
they are less important than the disc-driven torque.
Only in the \emph{MHD-cool-all} run, they are
of comparable importance to the disc torque,
probably due to higher clump masses. 
In the \emph{HD} runs,
there are fewer clumps, 
hence clump-clump interactions are less frequent
despite of the masses being on the higher end.
We note that the clump-clump interactions occurring in the
simulations excite the eccentricity only very briefly, 
hence they do not lead to higher eccentricities on orbital timescales.

\item The disc torque impacting the clumps has an inner disc component 
which yields a positive contribution,
and an outer component
which  yields a negative contribution. This is true for 
most of the clumps, and resembles qualitatively the behaviour
of the torque in Type I migration, although in our case
the amplitude of the torque is not due to the resonant interaction
with the disc; rather it is dominated by a local contribution.

\item The torque contribution arising from
material on orbits of at most 
$2 r_\text{hill}$ away from the clump's orbit
is enough to essentially explain the orbital 
evolution of the clumps due to disc-driven torques
(neglecting clump-clump interactions).
This suggests the development of a new prescription
for migration of protoplanetary clumps  to be used in population synthesis models of disc instability planets 
(e.g. \citep{mueller-18, forgan-18, schib-23}).
This will be the subject
of future work.

\item The nature of disc-driven torques is the same in magnetized
and unmagnetized discs, being controlled by the gravitational effect
of the local gas flow. This is true despite the  significant difference
in clump masses in MHD versus HD runs, the former being 1-2 orders of magnitude less massive and thus migrating slower.

\item The timescales of migration for each clump
are faster than what would be expected from Type I
migration and also faster than the viscous timescale, 
ruling out Type II migration.
In confirmation of this,
we do not observe fully developed gaps along the clump's
orbits although the most massive clumps 
in the \emph{MHD cool all} case
seem to open partial gaps,
that do not extend around the full azimuth. Both the
large turbulent viscosity and the fast migration
timescale are likely responsible for the
stifling of gap opening.
Instead, the libration timescales are  comparable
or shorter than the migration timescale, consistent
with Type III migration.

\end{itemize}

The eccentricities 
of the clumps measured in our simulations
are 
$\approx 0.1$ and not much influenced
by clump-clump interactions over the long term.
This is even true for the two mergers
we observed in the \emph{MHD cool all} run.
They lead neither to a strong increase
in eccentricity nor in inclination
of the orbits.
This is probably because the respective clumps
were almost co-planar, with an angle
between their angular momentum vectors of
$\approx 1 \degree$.
\citet{matzkevich-24}  simulated 
colliding clumps that formed through disc instability, without
including the surrounding disc but exploring a wide parameter space
in terms of orbital configurations and clump masses.
Although they observed perfect mergers, whereby the mass of the 
new protoplanet corresponds to the sum of the colliding bodies
(which also roughly holds for our mergers),
such events were rare.
However, their masses were considerably larger than
ours, ranging from $3$--$10\,M_\text{jup}$;
while the upper end of the mass 
range was more stable against 
disruptions from collisions.
They found that while
even collisions with
impact velocity $v_\text{imp}$ below the 
mutual escape velocity $v_\text{esc}$ 
of the clumps
could result in substantial mass
loss or trigger dynamic collapse,
$v_\text{imp} \lesssim v_\text{esc}$
was found to be a necessary condition for perfect 
merging.
This condition is also fulfilled in the 
(perfect) mergers
observed in our simulations.

The consistency in the nature of the disc-driven torque between the
MHD and the HD runs is not a trivial result since the former have
been shown to be significantly more turbulent, having an effective
``alpha'' viscosity as high as $0.2-0.4$ due to a combination of the
gravitational and Maxwell stress 
(see \citet{deng-20}). Turbulent
diffusion is known to play a role in the saturation of the horseshoe
drag in non-self-gravitating discs  \citep{BaruteauLin2010}.
Equivalently, here turbulent
diffusion could impact the amplitude of the dominant local torque
component by maintaining a sufficiently high mass flow near the
Hill sphere. In that respect, the HD and MHD runs might behave
similarly because the effective viscosity is different between the
two but only by factors of 2-3.

A high mass diffusion rate through the Hill sphere
likely plays a key role in causing a Type III-like migration behaviour by
keeping the co-orbital torque unsaturated.
In conventional Type III
migration studies, which are carried out in non-turbulent discs,
it
has been shown that the fastest form of Type III migration, 
runaway migration,
does not occur,
rather the acceleration saturates after an initial increase 
\citep{peplinski-doctoral}.
Here we have shown 
(see section \ref{ch-discussion}) that the
acceleration does not have a monotonic evolution,
rather it is highly fluctuating.
This underpins how the relevant regime is only superficially akin to the Type III migration mode described in the literature. Instead, our results suggest that our 
migration regime,
even in absence of clump-clump interaction,
is highly stochastic.

The stochasticity of migration observed in our simulations
reflects the gravito-magneto-turbulent regime of the flow
(with only gravito-turbulence in absence of the magnetic field).
In \citet{lin-24-chaotic}, who studied
migration of low-mass planets, it was  found
that for a highly turbulent flow,
chaotic migration dominates over 
classical Type I migration.
They determined a critical transition mass
for the planet to undergo chaotic migration,
around $q \lesssim \lambda \alpha h^3$
where $q$ is the planet to star 
mass ratio.
The factor $\lambda$ was determined
to be
$4 < \lambda < 25$ \citep{lin-24-chaotic}.
In the case of our discs, 
the critical limit would be
in the range of 
$0.05 < q_\text{crit} < 0.3$.
Therefore, chaotic migration
is compatible with the conditions
in our simulations.

Our simulations allow to study the effect of clump-clump interactions
but
we did not observe any ejections of
clumps. Such ejections have been previously observed
in simulations with sink particles
\citep{boss-23} and 
could be related to the 
free-floating planets, observed e.g. 
in \citet{miret-roig-22}.
It is important to realize that the ineffectiveness of clump-clump collisions in scattering protoplanets out of the disc, a major result of
our work, might be in part due to the limited spatial resolution. 
Indeed, since the clump sizes can not shrink below the
gravitational softening of 0.05 AU, collisions remain forcefully soft, which
could underestimate the exchange of kinetic energy and angular momentum
between clumps, 
thereby suppressing ejections artificially. 
However, previous
one-dimensional collapse models
as well as 
3-dimensional  hydrodynamical simulations of individual collapsing clumps 
\citep{helled-08, galvagni-12, helled-14}
have computed characteristic pre-collapse timescales,
defined as the time required for the clump to become dense
and hot enough to reach molecular hydrogen dissociation,
after which a much faster
dynamical collapse to planetary sizes would follow.
Pre-collapse
timescales range from a few $1000$ to a few $10^5\,\text{yr}$ for clumps of 1-10 Jupiter masses with solar metallicity 
and filled with interstellar-size grains,
with the shortest timescales 
holding for the upper end of this mass range.
The clumps in our simulation reach a maximal mass
of the order of a Jupiter mass, 
but some may grow
further in the \emph{MHD cool all} case.
If that is the case, they would collapse
much faster than the lifetime of the disc,
leading to a later evolutionary phase where clump-clump
interactions could become much stronger due to the 
smaller sizes of the clumps---a regime we have not covered 
in our simulations.
However, for the Jupiter-mass clumps, which we actually
observe in our simulation,
the pre-collapse timescales are comparable to the disc lifetime,
hence clump-clump collisions would be as soft as modelled in our simulations.
In the \emph{MHD cool disc} run,
which is probably more realistic as it 
accounts for the clumps' interiors being 
optically thick
(see section \ref{ch-setup}),
the clumps do not reach Jupiter masses.
For the Neptune-mass clumps
observed in our magnetized runs 
an even longer pre-collapse phase
is expected.
In further support of this statement, preliminary work using 1D collapse calculations similar to those of
\citet{helled-08}
but starting  with the clump's density and temperature occurring in our simulations
(Kubli et al., in prep.),
show that collapse timescales typically exceed a few $10^6$ yr for clumps below Saturn masses, even for higher than solar metallicities.

We can estimate directly the effect of clump compactness on
the relative velocity of the clumps. 
As mentioned above, the mutual gravitational acceleration of the two interacting clumps will increase with decreasing
impact parameter,
and more compact clumps can encounter one another
with smaller impact parameters.
We can estimate
the impact parameter
below which the relative velocity of the clumps 
becomes comparable to the Keplerian orbital velocity.
Computing $d_\text{min} = 2G(m_1+m_2)/v_\text{Kepler}^2$
yields values of $40$ -- $400$  times smaller
(for $0.5M_\text{jup}$ -- $0.01 M_\text{jup}$) than
the sizes of our clumps.
This implies that protoplanets
would have to collapse by more than an order of magnitude in order
for ejections to occur in the inner disc region,
where the clumps form 
in our simulations.
Due to the long collapse timescales,
such a compact configuration is only relevant at a much later evolutionary stage compared to the timescales probed
by our simulations, confirming the robustness of our findings.

A caveat is that  additional mechanisms such as pre-enrichment,
grain growth to pebble sizes
and grain sedimentation,
would all concur to reduce opacities,
especially at metallicities higher than solar.
\citet{helled-11}
have shown that the opacity reduction can decrease the pre-collapse timescales
by up to two orders of magnitude. Nevertheless, even with such an acceleration of the pre-collapse stage, the Neptune-sized clumps in our MHD simulations would still evolve slowly enough to remain diffuse and undergo soft collisions over the short timescales probed by our simulations.
The interplay between contraction timescales and dynamics will have to be investigated in the future. As already mentioned,
\citet{matzkevich-24} found that 
clump-clump collisions can trigger fast collapse under certain conditions, especially for older, evolved clumps. 
In the latter case though,
the collapse would occur after perfect merging,
hence no ejection would occur.

The inability of clump-clump collisions to eject protoplanets
can also depend on where in the disc these collisions occur.
Our discs
are relatively compact,
hence by construction there are no encounters
occurring beyond 20-25 AU.
We can then ask at what distance from the centre,
given a more extended disc,
clump-clump collisions with properties
as those in our simulations,
would lead to ejection.
This will happen
when the kinetic energy transferred in the collision becomes larger
than the clump's binding energy to the central star.
Equivalently, one can compare the relative velocity in the collision
with the local Keplerian velocity.
For this purpose we measured the relative velocity of clump pairs undergoing a close encounter.
In the case presented in fig. \ref{fig-clump-clump}, when they are 
at closest distance to each other, their relative velocity is
$0.35\,\text{AU}/\text{yr}$, whereas the Keplerian velocity at their
position is $2.12\,\text{AU}/\text{yr}$.
Similar values are found for other close encounters.
The measured relative velocity would correspond to the Keplerian velocity at a radial distance of $\approx 300\,\text{AU}$ for the same disc
mass and stellar mass. Of course this distance could be somewhat smaller
in discs around lower mass stars, but only by a small factor.

Overall, our findings in the simulations combined with the arguments just outlined lead us to conclude that ejections should be a rare phenomenon in protoplanetary discs undergoing fragmentation,
thereby not yielding a significant contribution to the population of free-floating planets.
These could be produced via other mechanisms,
involving still some
form of gravitational collapse in the gas phase triggered by other phenomena,
for example as a result of tidal interactions of protoplanetary discs
\citep{fu-freefloat-25}.
Instead, fast, stochastic migration is the distinctive phenomenon governing
the evolution of orbits of clumps arising in disc instability.

\section*{Acknowledgements}
We thank Christian Reinhardt for 
interesting discussions and helpful comments.
This work is supported by the Swiss Platform for Advanced Scientific Computing (PASC) project SPH-EXA2 and by NCCR PlanetS.
We also thank the Swiss National Supercomputing Center (CSCS)  where the simulations were carried out on the
PizDaint and Eiger supercomputers on the "uzh3" rolling account.

\section*{Data Availability}
The data files that support our analysis will be made available upon reasonable request.
 



\bibliographystyle{mnras}
\bibliography{quellen} 


\newpage
\appendix
\section{Additional figures}

\begin{figure}
    \centering
    \includegraphics[width=\columnwidth]{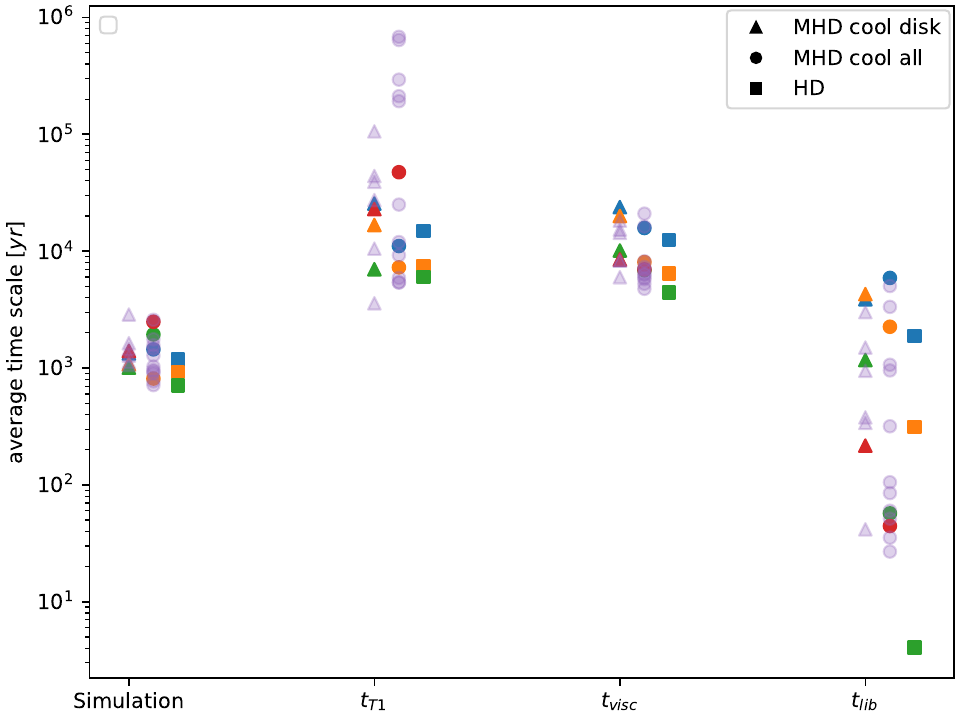}
    \caption{Comparison of different time scales
    relative to migration time (absolute times of fig. \ref{fig-timescales-rel}).}
    \label{fig-timescales-abs}
\end{figure}
\begin{figure*}
    \includegraphics[width=\textwidth]{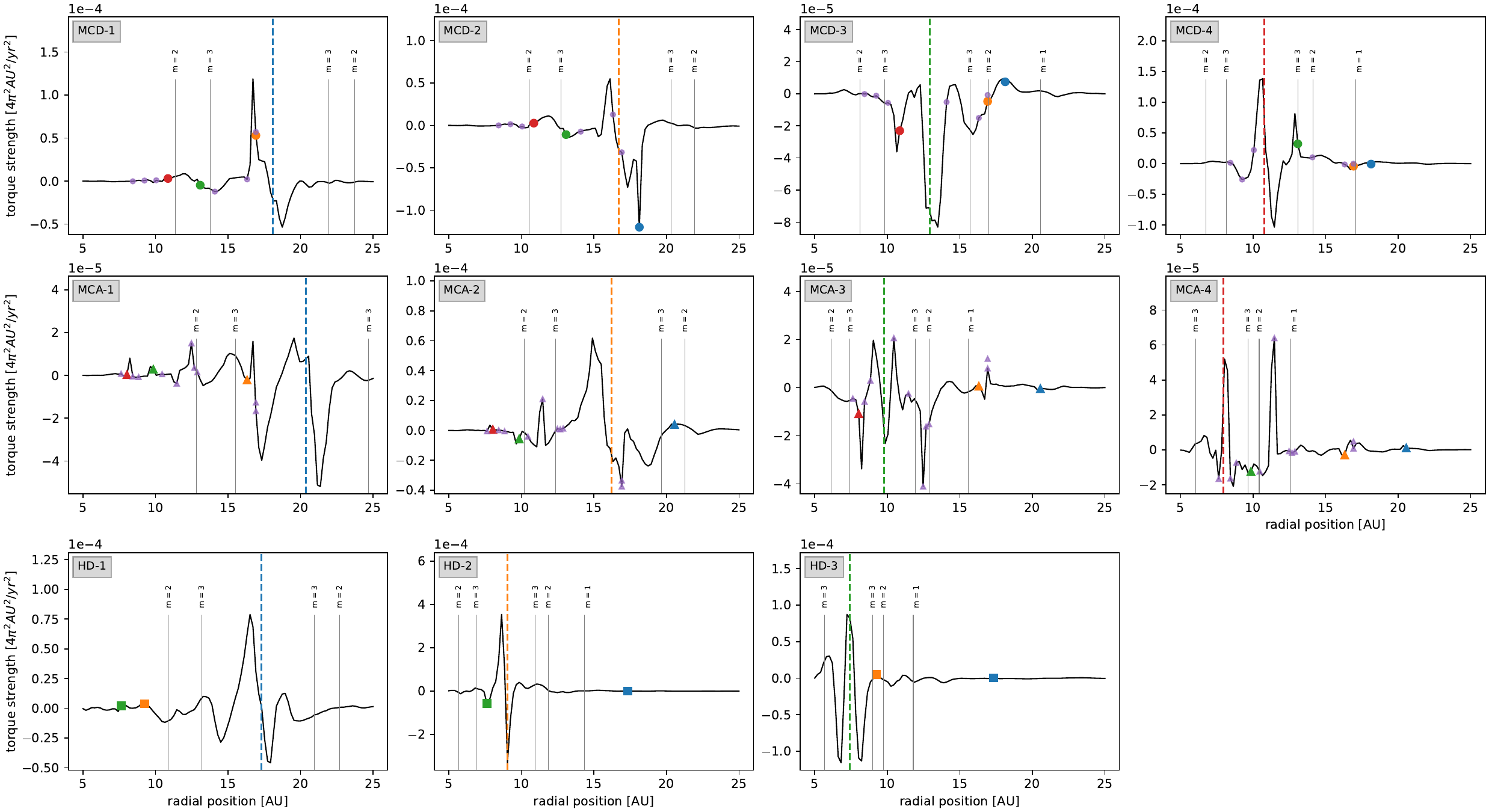}
    \caption{Same as 
    fig.~\ref{fig-torque-profiles} but
    at $t \sim \SI{199}{yr}$.
    The shape of the profile
    changes fast (e.g. 
    the spike to the interior of MCA-1 that exists in fig. \ref{fig-torque-profiles} has disappeared after half an orbit).}    \label{fig-torque-profiles125}
\end{figure*}


\bsp	
\label{lastpage}

\end{document}